\documentclass{jafm}  
\usepackage[T1]{fontenc}		% Silbentrennung auch bei Wörtern mit Umlaut
\usepackage[utf8]{inputenc}	% LaTeX erkennt Umlaute --> Eingabe ä statt \"a möglich
\usepackage{amsmath}
\usepackage{amsfonts}
\usepackage{amssymb}
\usepackage{graphicx}
\usepackage{hyperref}   % use for hypertext links, including those to external documents and URLs
\usepackage{amsmath}
\usepackage{amsfonts}
\usepackage{amssymb}
\usepackage{pifont}
\usepackage[multidot]{grffile}
\usepackage{framed}
\newcommand*{\factor}{0.97}			%adapt figure width when changing column number
\usepackage{enumerate}
\usepackage{hyperref}
\usepackage{float}
\usepackage{placeins}

\usepackage{hanging}    %for reference section

\renewcommand{\thesection}{\arabic{section}}

\newenvironment{myfont}{\fontfamily{phv}\selectfont}{\par}

\received_date{------}
\accepted_date{------}
\begin{document}
\title{Analytical and numerical investigation of the airflow in face masks used for protection against COVID-19 virus -- implications for mask design and usage }
\author{
R. Perić\inst{1}$^{\dagger}$ \and 
M. Perić\inst{2}  }
\institute{ Institute for Fluid Dynamics and Ship Theory, Hamburg University of Technology, Am Schwarzenberg-Campus 4 (C), 21073 Hamburg, Germany \and Institute of Ship Technology, Ocean Engineering and Transport Systems, Faculty of Engineering, University of Duisburg-Essen, Bismarckstraße 69, 47057, Duisburg}
\email{robinson.peric@tuhh.de}
\abstract{The use of face masks for the general public has been suggested in literature as a means to decrease virus transmission during the global COVID-19 pandemic. However, literature findings indicate that most mask designs do not provide reliable protection. This paper investigates the hypothesis that the impaired protection is mainly due to imperfect fitting of the masks, so that airflow, which contains virus-transporting droplets, can leak through gaps into or out of the mask. The fluid dynamics of face masks are investigated via analytical and numerical computations. The results demonstrate that the flow can be satisfactorily predicted by simplified analytical 1D-flow models, by efficient 2D-flow simulations and by 3D-flow simulations. The present results show that already gap heights larger than $0.1\, \mathrm{mm}$ can result in the mask not fulfilling FFP2 or FFP3 standards, and for gap heights of ca. $1\, \mathrm{mm}$ most of the airflow and droplets may pass through the gap. The implications of these findings are discussed and improvements to existing mask designs are suggested.
}
\maketitle

\section{Introduction}
\label{SECintro}
During the virus pandemic COVID-19 in 2020, several countries have either enforced or discussed whether wearing face masks should be compulsory in public places (Howard, 2020). 
Opposing arguments  include that masks should be reserved for healthcare workers and that the use  of  medical masks  in  the  community  may create  a  false   sense   of   security (World Health Organization, 2020).
However, the global shortage of masks may not persist indefinitely so that protective masks may become available to the public in  future.
Arguments in favor of using face masks for the general public include indications by several studies that face masks can reduce viral exposure and infection risks (e.g. van der Sande et al., 2008). Protective effects of masks were demonstrated for various  severe infections such as SARS, tuberculosis and pandemic influenza (e.g. Andersen, 2019; Jung et al., 2014, and references therein) and also for dust, oil, or combustion-exhaust particles (Penconek et al., 2012; Ntlailane and Wichmann, 2019).

Virus transmission can occur via  direct  contact  with secretions and via  exhaled water droplets  (Tellier, 2006). The former risk can be reduced by hygiene  practices, such as hand washing and not touching one's face; furthermore, the virus was found to survive at most several days on surfaces (Kampf et al., 2020). However, reducing the risk of virus transmission via droplets is complicated by the different particle sizes, which roughly range from larger droplets in the order of $100 \, \mathrm{\mu m}$ to smaller aerosol-size droplets in the order of $<1\, \mathrm{\mu m}$ (Tellier, 2006; Fabian et al., 2008). Larger particles ($>20  \, \mathrm{\mu m}$) have shorter settling times, i.e. they fall to the ground within seconds up to a few minutes. Smaller particles, however, may float in the air for hours and  particles with a diameter $<3  \, \mathrm{\mu m}$ essentially do not settle (Tellier, 2006). Thus virus-contaminated droplets may accumulate in closed rooms such as workplaces or public transportation.  Although at present it is not known how long the virus remains infectious, results by van Doremalen et al. (2020) indicate that the virus could survive several hours in aerosols-size droplets, and Asadi et al. (2020) conclude from their literature review that even the smallest aerosol-size droplets might be able to transmit the virus during face-to-face conversation with an asymptomatic infected individual.

Fabian et al. (2008) found that humans exhale  $>500$ droplets per liter of air, of which $99.9\%$ had diameters  between $0.3\, \mathrm{\mu m}$ and $ 5 \, \mathrm{\mu m} $, and $87\%$ were smaller than $ 1\, \mathrm{\mu m}$. That most exhaled particles have diameters below $1\, \mathrm{\mu m}$ has been confirmed by other authors (e.g. Fairchild and Stampfer, 1987; Papineni and Rosenthal, 1997; Edwards et al., 2004). Thus this work focuses on  particles smaller than $ 5 \, \mathrm{\mu m} $. 

To reduce the risk of inhaling such droplets, self-made-masks, surgical masks and so-called respirators, such as Filtering Face Piece (FFP) masks, are under discussion (cf. Steinle et al., 2018). 

The performance of face masks is typically assessed via experiments that determine the \textit{filtration efficiency} (FE) and the \textit{total inward leakage} (TIL). The \textit{filtration efficiency} is the percentage of particles that do not pass through the filter if the mask is tightly fitted. For example, Mueller et al. (2018) reported average filtration efficiencies of handkerchiefs ($\mathrm{FE}=22.7\%$), T-shirts ($\mathrm{FE}=42.5\%$), surgical masks ($66.2\% \leq \mathrm{FE} \leq 88.7\%$) and FFP3-masks ($\mathrm{FE}=99.3\%$, corresponding to the mask in Fig. \ref{FIGffp3mask}). Thus, whereas typical household fabrics may not provide sufficient protection, industrial materials from which FFP3-masks are made typically have satisfactory filtering qualities. 

However, the filtration efficiency of the filter material is not sufficient to assess the protection offered by a mask; rather, the protection  can be measured by the \textit{total inward leakage}, which is the percentage of particles that enter the mask through both the filter and the face-seal leakage. Alternatively, the protection factor (PF) can be used, for which holds  $\mathrm{PF} = 1/\mathrm{TIL}$ (van der Sande et al., 2008). 

Milton et al. (2013) found that surgical masks reduced the number of influenza virus droplet that were emitted by ca. $75\%$ compared to test persons not wearing masks if the droplets were larger than $5\, \mathrm{\mu m}$, but for smaller droplets the surgical masks provided no substantial protection.

Therefore, in the following the focus will be only on FFP-masks, which are typically more effective than surgical or home-made masks (van der Sande et al., 2008). FFP-masks can be subdivided in classes FFP1 to FFP3 with filtration efficiencies of $80\%$ (FFP1), $94\%$ (FFP2) and $99\%$ (FFP3), respectively, and leakage rates of less than $22\%$ (FFP1), $8\%$ (FFP2) and $2\%$ (FFP3) (cf. Seidler et al., 2004; Lee et al., 2016; European norm EN 149:2001+A1:2009). Thus technically it would appear that FFP3-masks provide satisfactory protection, and indeed they are widely used in hospitals. As outlined in the following, though, literature suggests that many FFP3-masks do not provide reliable protection, especially if not properly fitted.

Steinle et al. (2018) investigated FFP2-masks with median filtration efficiency of $\mathrm{FE}\geq 98\%$ and found that the total inward leakage varied between $0\% \leq \mathrm{TIL} \leq 84.4\%$. 
Cherrie et al. (2018), Lee et al. (2016) and Jung (2014) reported similar values and argued that, regardless of the quality of the filter materials, FFP-type masks may not provide reliable protection if the mask does not fit tightly. 
Lee et al. (2017) reported face-seal leakage surrounding the chin and the cheeks for FFP-masks, because the masks did not fit all wearers. Furthermore, they observed that the percentage of particles that penetrated through face-seal leaks increased for low air intake and for particle diameters below $0.1 \, \mathrm{\mu m}$.
Children were found to be less protected by FFP-type masks, which was attributed to inferior fitting of the masks on smaller faces (van der Sande et al., 2008). 

Moreover, filtration efficiencies for masks may be different depending on the particles used. Penconek et al. (2013) found that commercially available FFP2- and FFP3-masks did not provide sufficient protection against diesel exhaust fumes. Filtration efficiency is typically tested with NaCl particles or paraffin oil droplets (Penconek et al., 2013), but water droplets appear to be rarely used.

Performing activities such as exercising, nodding or shaking was found to affect total inward leakage, although the total inward leakage varied mostly by factor $2$ or less (van der Sande et al., 2008).

However, how tightly does a mask have to fit and how large may gaps between mask and face be before the mask ceases to provide the promised protection? The aim of the present work is to answer these questions via analytical and numerical flow computations for a generic mask. From these findings, recommendations for the design of more effective  industrial and self-made masks are derived, with focus on re-usability of the masks.

\section{Theory}
\label{SECtheory}
\begin{figure}[H]
\begin{center}
\includegraphics[width=0.9\linewidth]{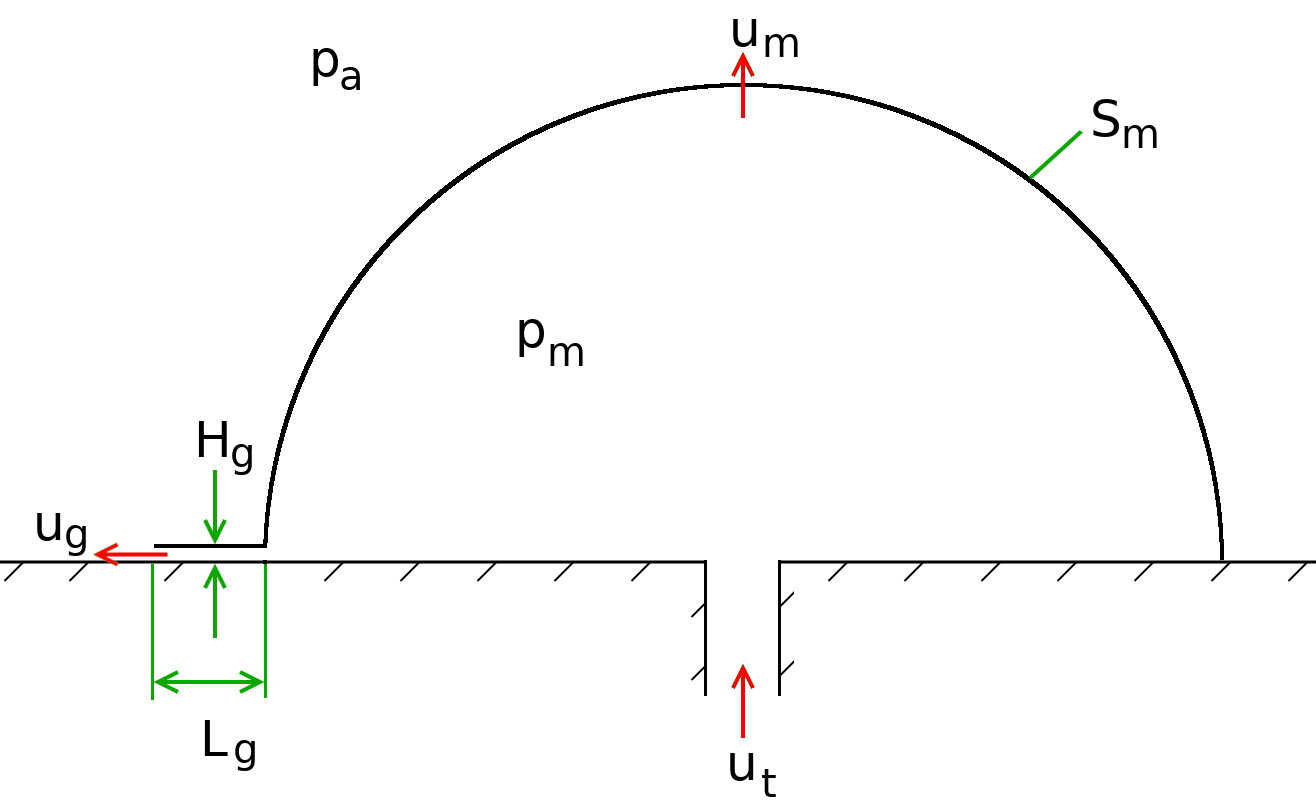}
\caption{Simplified geometry for airflow through a generic face mask, which has a gap with height $H_{\mathrm{g}}$ and length $L_{\mathrm{g}}$ over a width  $B_{\mathrm{g}}$ along its perimeter and is otherwise tightly fitted to the face; average velocities $u_{\mathrm{t}}$, $u_{\mathrm{m}}$ and $u_{\mathrm{g}}$ denote airflow through  nose cross-sectional area $S_{\mathrm{t}}$,  mask filter surface $S_{\mathrm{m}}$ and gap cross-sectional area $S_{\mathrm{g}}$; $p_{\mathrm{m}}$ and $p_{\mathrm{a}}$ denote the pressure inside and outside of the mask} \label{FIG2DtheoryGeom}
\end{center}
\end{figure}

To investigate the airflow through face masks, the  problem can be reduced to one-dimensional (1D) flow for a simplified geometry (cf. Fig. \ref{FIG2DtheoryGeom}) by the following assumptions. The fluid is considered as incompressible because the Mach number $Ma$ is well below $0.3$ (cf. Ferziger et al., 2020). The pressure within the mask is assumed to be uniform, so $\partial p/ \partial x_{i} \approx  0$ for all directions $x_{i}$; thus the surface area $S_{\mathrm{m}}$ of the mask filter-piece influences  the flow through the filter, but  the geometrical shape of the mask has a negligible influence. Therefore, the mask geometry can be simplified to a half-sphere e.g. with radius $r\approx 0.0502\, \mathrm{m}$ and surface area $S_{\mathrm{m}}\approx 0.015833\, \mathrm{m^{2}}$.   
The mask has a uniform gap along a width $B_{\mathrm{g}}$ of its perimeter and otherwise fits tightly to the face. The mask rim has length $L_{\mathrm{g}}$ with constant gap height $H_{\mathrm{g}}$. 
Face and nose are approximated as a plane pierced by a channel with cross-sectional area $S_{\mathrm{t}}$ (cf. Fig. \ref{FIG2DtheoryGeom}). No flow occurs through the face, the nose or the boundaries of the gap. 

When inhaling or exhaling, the total (volumetric) flow rate through the nose is $F_{\mathrm{t}}=u_{\mathrm{t}}S_{t}$ with average velocity $u_{\mathrm{t}}$ and cross-sectional area $S_{\mathrm{t}}$. From mass conservation follows 
\begin{equation}
F_{\mathrm{t}} = F_{\mathrm{g}} + F_{\mathrm{m}}  \quad ,
\label{EQFtFgFm}
\end{equation}
with flow rate through the gap $F_{\mathrm{g}} = u_{\mathrm{g}}S_{\mathrm{g}}$, average flow velocity $u_{\mathrm{g}}$ within the gap, gap cross-sectional area $S_{\mathrm{g}}$, flow rate through the mask filter $F_{\mathrm{m}} = u_{\mathrm{m}}S_{\mathrm{m}}$, average flow velocity $u_{\mathrm{m}}$ through the surface $S_{\mathrm{m}}$ of the masks filter-piece.

Within the gap, the flow is assumed to be fully-developed laminar Pouseuille flow, which is justified because the average  gap velocities $u_{\mathrm{g}}$ within the framework of this analysis were below the  critical velocity 
\begin{equation}
u_{\mathrm{crit}} \approx \frac{\nu \mathrm{Re_{crit}}}{D_{\mathrm{h}}}  \quad ,
\label{EQRecrit}
\end{equation}

with hydraulic diameter $D_{\mathrm{h}}=2H_{\mathrm{g}}$, gap heights $H_{\mathrm{g}} \in [0.1\, \mathrm{mm},1\, \mathrm{mm}]$, kinematic viscosity $\nu \approx 1.5 \cdot 10^{-5} \, \mathrm{m^{2}/s} $ for air at $20\, \mathrm{C^{\circ}}$ room temperature and critical Reynolds number for plane channel flow $\mathrm{Re_{crit}} \approx 3000$   (cf. Schlichting and Gersten, 2017, p. 104). Gap Reynolds numbers in this work were within $\mathrm{Re} \in [5, 1800]$.

Pressure losses occur in the flow at the gap inlet, within the gap, at the gap outlet, and at the mask filter-piece.
The combined pressure loss at  gap inlet and gap outlet can be expressed as
\begin{equation}
\Delta p_{\mathrm{g,io}} = \frac{u_{\mathrm{t}}}{|u_{\mathrm{t}}|} \zeta  \frac{\rho }{2}  u_{\mathrm{g}}^{2}  \quad ,
\label{EQpg1}
\end{equation}
with loss coefficient $\zeta = \zeta_{\mathrm{in}} + \zeta_{\mathrm{out}}$, density of air $\rho\approx 1.2\, \mathrm{kg/m^{3}}$, average gap velocity $u_{\mathrm{g}}$, average flow velocity $u_{\mathrm{t}}$ through the nose and its absolute value $|u_{\mathrm{t}}|=\sqrt{u_{\mathrm{t}}^{2}}$; the term $u_{\mathrm{t}}/|u_{\mathrm{t}}|$ is included to obtain the correct sign for both inhaling and exhaling. 

The loss coefficients $\zeta_{\mathrm{in}}$ and  $\zeta_{\mathrm{out}}$ were taken from Idelchik (1986, p. 92 and p. 128) as $\zeta_{\mathrm{in}} = 0.5$ and $\zeta_{\mathrm{out}} = 1.0$. 

Assuming a fully-developed laminar flow in a plane-channel, the pressure loss within the gap is 
\begin{equation}
\Delta p_{\mathrm{g,2}} = \frac{12 \mu L_{\mathrm{g}}}{H_{\mathrm{g}}^{2}}  u_{\mathrm{g}}\quad ,
\label{EQpg2}
\end{equation}
with dynamic viscosity of air $\mu \approx 1.8 \cdot 10^{-5} \, \mathrm{Pa s}$, gap length $L_{\mathrm{g}}$, gap height $H_{\mathrm{g}}$ and average gap velocity $u_{\mathrm{g}}$.

The pressure loss due to the flow through the mask surface $S_{\mathrm{m}}$ is approximately 

\begin{equation}
\Delta p = p_{\mathrm{m}} - p_{a} \approx C_{\mathrm{m}} \rho u_{\mathrm{m}} \quad ,
\label{EQpm}
\end{equation}
with viscous porous resistance $C_{\mathrm{m}}$, which is a property of the mask filter material, density $\rho$ of air and velocity $u_{\mathrm{m}}$ of the airflow through the mask, which can be computed from the flow rate $F_{\mathrm{m}}$ through the mask. 

It is expected that the assumption, that the pressure loss depends linearly on the velocity, can be made with good approximation for these flow rates. Unfortunately, only qualitative curves were available to the authors to validate this statement. However, Jung et al. (2014) present experimental data for pressure drop $\Delta p$ for flow rate $F_{\mathrm{t}}=30\, \mathrm{L/min}$ and $F_{\mathrm{t}}=85\, \mathrm{L/min}$ for FFP-type masks, and even when linearly scaling the pressure drop $\Delta p(30\, \mathrm{L/min})$ for  $F_{\mathrm{t}}=30\, \mathrm{L/min}$ via $\Delta p(85\, \mathrm{L/min}) \approx \Delta p(30\, \mathrm{L/min}) \frac{85\, \mathrm{L/min}}{30\, \mathrm{L/min}}$ gives only differences to the actually measured value of $15\%$ to $40\%$. This is acceptable for the present purposes, because the  results in Sect. \ref{SECres} show that changing $\Delta p$ by even $300\%$ changed the ratio $F_{\mathrm{g}}/F_{\mathrm{t}}$, i.e. gap flow rate divided by total flow rate, by ca. $10\%$ or less, i.e. the flow did not change qualitatively.

The pressure drop $\Delta p $ through the gap must be equal to the pressure drop through the mask filter-piece

\begin{equation}
\frac{u_{\mathrm{t}}}{|u_{\mathrm{t}}|}  \zeta  \frac{\rho }{2} u_{\mathrm{g}}^{2} + \frac{12 \mu L_{\mathrm{g}}}{H_{\mathrm{g}}^{2}}  u_{\mathrm{g}} = C_{\mathrm{m}} \rho u_{\mathrm{m}} \quad ,
\label{EQdpsum1}
\end{equation}
 and because  the total flow rate  $F_{\mathrm{t}}=u_{\mathrm{t}}S_{\mathrm{t}}$ must equal the sum of the flow rates through mask $F_{\mathrm{m}}$ and gap $F_{\mathrm{g}}$  (cf. Eq. (\ref{EQFtFgFm})) 

\begin{equation}
F_{\mathrm{m}} = F_{\mathrm{t}} - F_{\mathrm{g}} = F_{t} - u_{\mathrm{g}} H_{\mathrm{g}} B_{\mathrm{g}}\quad ,
\label{EQflowratesum}
\end{equation}
and thus the flow velocity through the filter can be expressed as
\begin{equation}
 u_{\mathrm{m}} = \frac{F_{\mathrm{m}}}{S_{\mathrm{m}}} = \frac{ F_{t} - u_{\mathrm{g}} H_{\mathrm{g}} B_{\mathrm{g}}}{S_{\mathrm{m}}} \quad .
\label{EQvgfromflowratessum}
\end{equation}
Inserting Eq. (\ref{EQvgfromflowratessum}) into Eq. (\ref{EQdpsum1}) gives

\begin{equation}
\begin{aligned}
\frac{u_{\mathrm{t}}}{|u_{\mathrm{t}}|}  \zeta  \frac{\rho }{2} u_{\mathrm{g}}^{2} + \left( \frac{12 \mu L_{\mathrm{g}}}{H_{\mathrm{g}}^{2}} + \frac{C_{\mathrm{m}}\rho H_{\mathrm{g}}B_{\mathrm{g}}}{S_{\mathrm{m}}} \right) u_{\mathrm{g}} \\ - \frac{C_{\mathrm{m}} \rho F_{\mathrm{t}}}{S_{\mathrm{m}}}    = 0 \quad .
\label{EQdpfull}
\end{aligned}
\end{equation}

The solution for gap velocity $u_{\mathrm{g}}$ can be determined as follows:
\begin{equation}
a  u_{\mathrm{g}}^{2} + b  u_{\mathrm{g}} + c = 0 \quad ,
\label{EQdpsumsol1}
\end{equation}
\begin{equation}
a   = \frac{u_{\mathrm{t}}}{|u_{\mathrm{t}}|}  \frac{\zeta \rho}{2} \quad ,
\label{EQdpsumsol2}
\end{equation}
\begin{equation}
b = \frac{12 \mu L_{\mathrm{g}}}{H_{\mathrm{g}}^{2}} + \frac{C_{\mathrm{m}}\rho H_{\mathrm{g}}B_{\mathrm{g}}}{S_{\mathrm{m}}} \quad ,
\label{EQdpsumsol3}
\end{equation}

\begin{equation}
c = - \frac{C_{\mathrm{m}}\rho F_{\mathrm{t}}}{S_{\mathrm{m}}} \quad ,
\label{EQdpsumsol4}
\end{equation}

\begin{equation}
u_{\mathrm{g}} = \frac{-b + \sqrt{b^{2}-4ac}}{2a}\quad .
\label{EQvgsol}
\end{equation}

The average flow velocity through the mask $u_{\mathrm{m}}$, the pressure drop $\Delta p$ and the flow rates through the mask filter and through the gap can then be computed as
\begin{equation}
F_{\mathrm{m}} = u_{\mathrm{m}} S_{\mathrm{m}} \quad ,
\label{EQFmsol}
\end{equation}

\begin{equation}
\Delta p = C_{\mathrm{m}} \rho u_{\mathrm{m}} \quad ,
\label{EQdpsol}
\end{equation}

\begin{equation}
F_{\mathrm{g}} = u_{\mathrm{g}} H_{\mathrm{g}} B_{\mathrm{g}} \quad .
\label{EQ}
\end{equation}

This 1D-model allows the computation of the gap flow rate $F_{\mathrm{g}}$ as a function of gap width $B_{\mathrm{g}}$, gap height $H_{\mathrm{g}}$ and gap length $L_{\mathrm{g}}$.

\section{Simulation Setup}
\label{SECsetup}
Except for the largest particles, most  droplets exhaled during  breathing are in the range of typical tracer particles used in optical flow measurement techniques. For example, typical droplet diameters used in gas flows lie within $0.5\, \mathrm{\mu m}$ to $5\, \mathrm{\mu m}$ (cf. Tropea and Yarin, 2007). Thus,  the droplets of interest in this work can be assumed to follow the airflow and need not be resolved in the simulations. 

In order to validate the analytical model from Sect. \ref{SECtheory}, simulations are performed for using a 2D-geometry and a commercial computational fluid dynamics (CFD) software.
The governing equations for the single-phase flow simulations are the equation for mass conservation and the three equations for momentum conservation, collectively called the Navier-Stokes equations: 
\begin{equation}
\frac{\partial }{\partial t} \int_{V} \rho \ \mathrm{d}V + \int_{S} \rho \textbf{u}  \cdot \textbf{n} \ \mathrm{d}S =  0  \quad ,
\label{EQcontiacc}
\end{equation}
\begin{align}
\frac{\partial }{\partial t}  \int_{V} \rho u_{i} \ \mathrm{d}V 
+ \int_{S} \rho u_{i} \textbf{u}  \cdot \textbf{n} \ \mathrm{d}S =  \nonumber \\ 
\int_{S} (\tau_{ij}\textbf{i}_{j} - p\textbf{i}_{i}) \cdot \textbf{n} \ \mathrm{d}S 
+ \int_{V} \rho \textbf{g} \cdot \mathbf{i}_{i} \ \mathrm{d}V \quad ,
\label{EQnavier_stokesacc}
\end{align}
with volume $V $ of control volume (CV) bounded by the closed surface $S$, fluid velocity vector \textbf{u}  with the Cartesian components $u_{i}$,  unit vector \textbf{n} normal to $S$ and pointing outwards, time $t$, pressure $p$, fluid density $\rho$, components $\tau_{ij}$ of the viscous stress tensor and  unit vector \textbf{i}$_{j}$ in direction $ x_{j} $. 
 The fluid, gaseous air, is considered incompressible. 
Selected flow simulations were repeated with compressible air following the ideal gas law  and  it was verified that compressibility effects were negligible for the investigated cases.

\begin{figure}[H]
\begin{center}
\includegraphics[width=\linewidth]{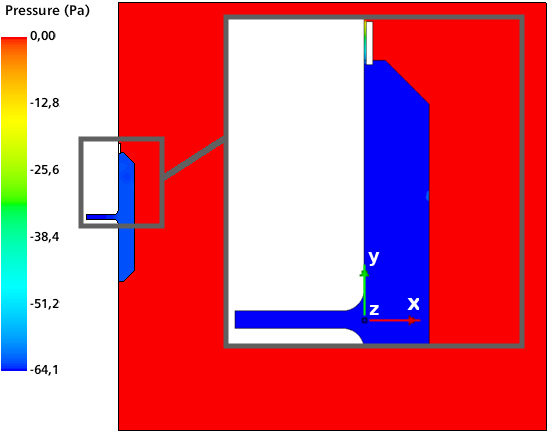} 
\caption{Computational domain for 2D-flow simulations through a generic mask showing the pressure distribution, with close-up of nose, mask and gap; note that throughout this work, the pressure $p$ is given relative to atmospheric pressure $p_{\infty}$; after the simulation is finished, the pressure for inhaling with flow rate $F_{\mathrm{t}}=30\, \mathrm{L/min}$ is nearly uniform inside the mask} \label{FIG2Ddom}
\end{center}
\end{figure}

The computational domain is box-shaped with  dimensions $0 \leq x \leq 0.4\, \mathrm{m}, -0.2\, \mathrm{m} \leq y \leq 0.2\, \mathrm{m}, 0 \leq z \leq 0.1\, \mathrm{m}$, with an extended channel that represents the nose as illustrated in Fig. \ref{FIG2Ddom}. 
The coordinate system has its origin in the center of the nose opening at the same level as the face, as shown in Fig. \ref{FIG2Ddom}, which is represented as a flat wall. As Fig. \ref{FIG2Ddom} demonstrates, the pressure within the mask is approximately uniform, therefore only the surface area $S_{\mathrm{m}}$,  not the geometrical shape of the mask, influences the results, so the mask was represented by a simplified polygonal shape. The mask surface is the same as in Sect. \ref{SECtheory}, $S_{\mathrm{m}}=0.015833\, \mathrm{m^{2}}$, and it is modeled as a porous interface with pressure drop $\Delta p $ according to Eq. (\ref{EQpm}), with viscous porous resistance $C_{\mathrm{m}}=2000\, \mathrm{m/s}$ unless stated otherwise. The top boundary ($y=0.2\, \mathrm{m}$) is set up as pressure outlet, with atmospheric pressure $p_{\infty}$ prescribed. Symmetry boundary conditions are prescribed at boundaries $x=0.4\, \mathrm{m}$, $y=-0.2\, \mathrm{m}$, $z=0\, \mathrm{m}$ and $z=0.1\, \mathrm{m}$. At the outward end surface $S_{\mathrm{t}}$ of the nose-channel, the  velocity $u_{\mathrm{t}}$ is prescribed so that $u_{\mathrm{t}}S_{\mathrm{t}}=F_{\mathrm{t}}$. All other boundaries are impermeable, no-slip wall boundaries. 

 The simulations were performed  quasi-two-dimensional (2D), i.e. the grid is only 1 cell thick in $z$-direction with symmetry conditions applied at front and back planes. %Thus the simulation is simplified to a plane channel-flow problem. 
 With correctly prescribed total flow rate $F_{\mathrm{t}}$, nose cross-section area  $S_{\mathrm{t}}$, mask filter surface $S_{\mathrm{m}}$ and gap cross-section area $S_{\mathrm{g}}$, equivalent results to the theoretical formulation from Sect. \ref{SECtheory} can be expected. As all gradients in $z$-direction are zero, the width $\Delta z$ of the domain can be chosen arbitrarily and the results can be scaled to the desired gap width $B_{\mathrm{g}}$ and mask filter surface $S_{\mathrm{m}}$. The basic configuration corresponds to a mask which fits tightly except over a width $B_{\mathrm{g}}=10\, \mathrm{cm}$ along  the perimeter. 
 
 To simulate the flow in a geometry with a shorter gap width, e.g. $B_{\mathrm{g}}=5\, \mathrm{cm}$, the simulation could be performed in three dimensions (3D) with a partially closed gap, which would increase the number of cells and the computational effort. In the following, a more efficient approach was selected so that the flow problem remains two-dimensional and the computational effort remains low. Consider that, if the gap were closed ($H_{\mathrm{g}}=0$), the flow rate through the mask filter  $F_{\mathrm{m}}=F_{\mathrm{t}}$, so the average velocity through the mask filter is $u_{\mathrm{m}}=u_{\mathrm{t}}S_{\mathrm{t}}/S_{\mathrm{m}}$. 
If the gap width $B_{\mathrm{g}}$ is halved,  the domain size in $z$-direction and thus $S_{\mathrm{t}}$ and $S_{\mathrm{m}}$ are halved as well. To maintain the same flow rate $F_{\mathrm{t}}$, $u_{\mathrm{t}}$ and thus also  $u_{\mathrm{m}}$ are doubled. To avoid that  Eq. (\ref{EQpm}) leads to twice the desired pressure drop along the mask, the larger mask filter is modeled by halving $C_{\mathrm{m}}$. Thus the same flow rate and pressure loss occur through the mask as previously. So if the reference gap width is as in the present case $B_{\mathrm{g, ref}}=10\, \mathrm{cm}$, then selecting $C_{\mathrm{m}}=C_{\mathrm{m,ref}} B_{\mathrm{g}}/B_{\mathrm{g,ref}}$ provides the desired solution for simulations with different gap widths.
 
\begin{figure}[H]
\begin{center}
\includegraphics[width=\linewidth]{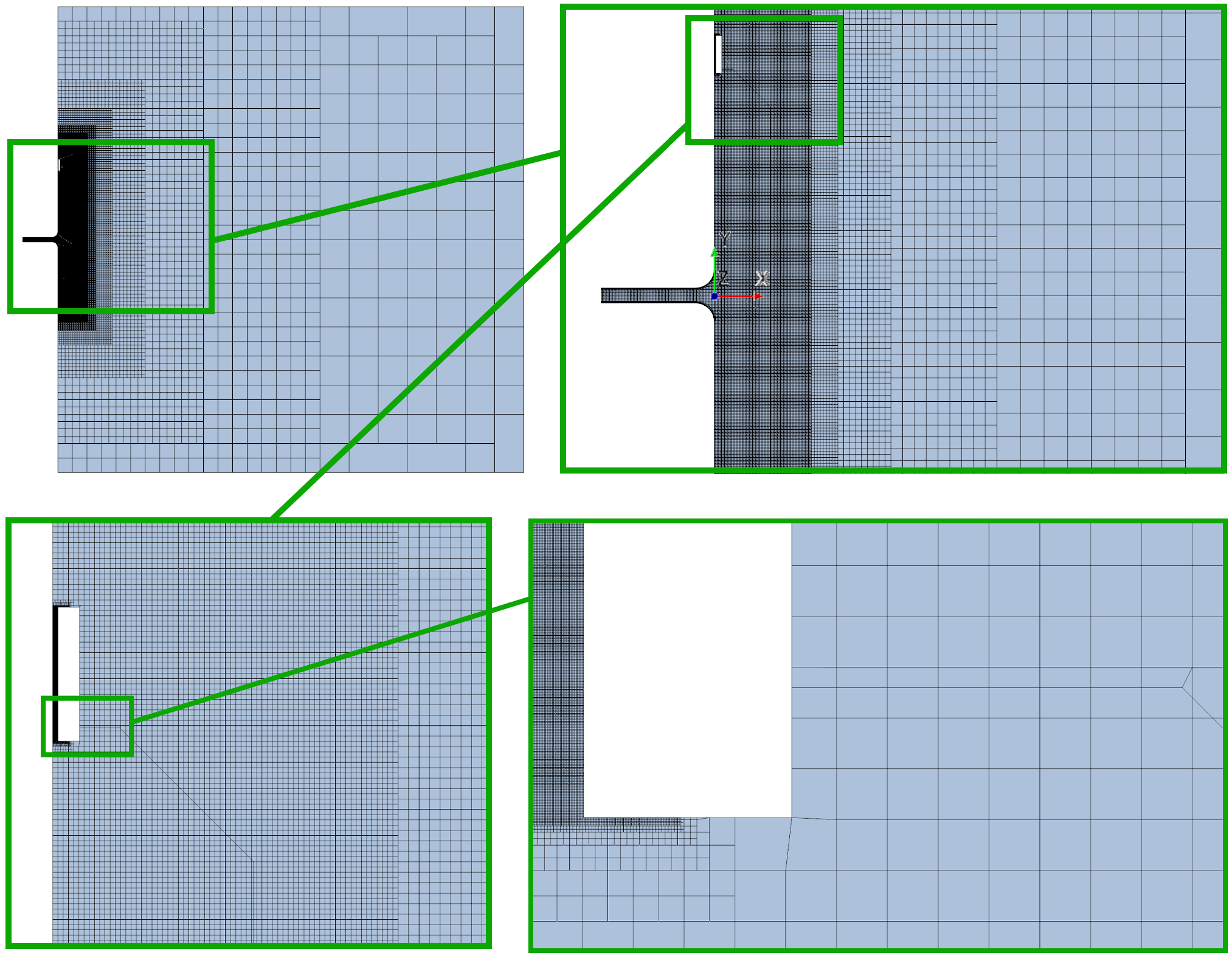} 
\caption{Computational grid with local  refinement, especially in the vicinity of the gap; full domain (top left) and close-up of nose and mask (top right), gap (bottom left) and gap entry (bottom right)} \label{FIG2mesh}
\end{center}
\end{figure}

The solution domain was discretized with a rectilinear grid with local mesh refinement as shown in Fig. \ref{FIG2mesh}, so that  the gap was resolved with at least $10$  cells per gap height $H_{\mathrm{g}}$. For the grid dependence study, the  grids were refined uniformly by halving the cells in $x$- and $y$-directions. Depending on the gap size, the grids consisted of ca. $10\, 000$ to $160\, 000$ cells.

All simulations were performed using the commercial flow solver Simcenter STAR-CCM+ (version 15.02.007-R8) by Siemens. The solver is based on the finite volume method (FVM) and the implicit unsteady segregated solver was used.
All approximations were of second order. The under-relaxation factors were $0.8$ for velocities and $0.2$ for pressure. The initial conditions were pressure $p=p_{\infty}$,  velocity $\mathbf{u}=0$ and density $\rho_{\mathrm{air}}=1.2\, \mathrm{kg/m^{3}}$. 
The time step was $\Delta t = 0.004\, \mathrm{s}$ and  $6$ outer  iterations were performed per time step. The total simulated time was selected $t_{\mathrm{end}}=0.5\, \mathrm{s}$, at which point a converged, quasi-steady solution was obtained in all simulations. Detailed information on finite-volume-based flow simulations can be found e.g. in Ferziger et al. (2020).

Simulations were performed for inhaling of air and compared to theoretical predictions from Sect. \ref{SECtheory} for different gap height $H_{\mathrm{g}} \in [0.1\, \mathrm{mm}, 1\, \mathrm{mm}]$, gap width $B_{\mathrm{g}} \in [2.5\, \mathrm{cm}, 10\, \mathrm{cm}]$ and  total flow rate $F_{\mathrm{t}} \in [30\, \mathrm{L/min}, 95\, \mathrm{L/min}]$.

To verify that the conclusions drawn from the  2D-flow simulations are applicable to realistic 3D-flow through face masks, also 3D-flow simulations were performed for a few representative configurations.  

\begin{figure}[H]
\begin{center}
\includegraphics[width=0.49\linewidth]{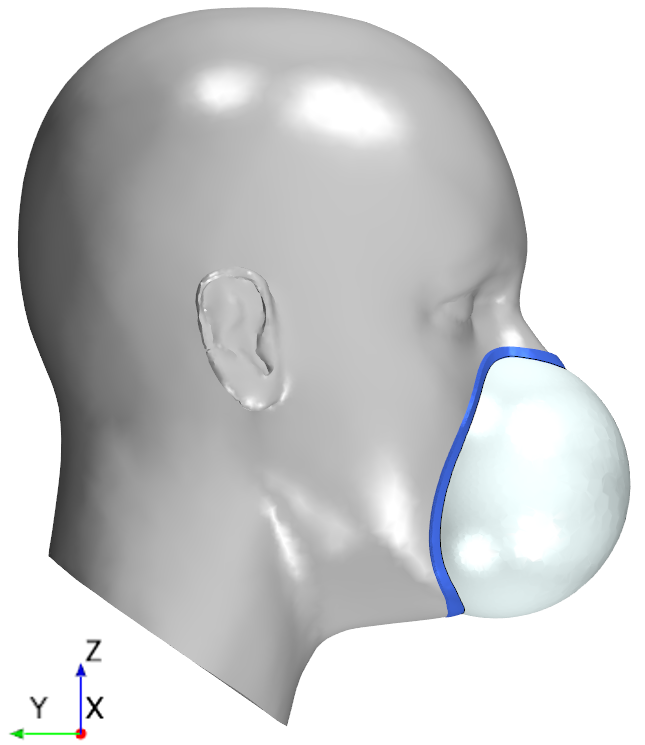}
\includegraphics[width=0.43\linewidth]{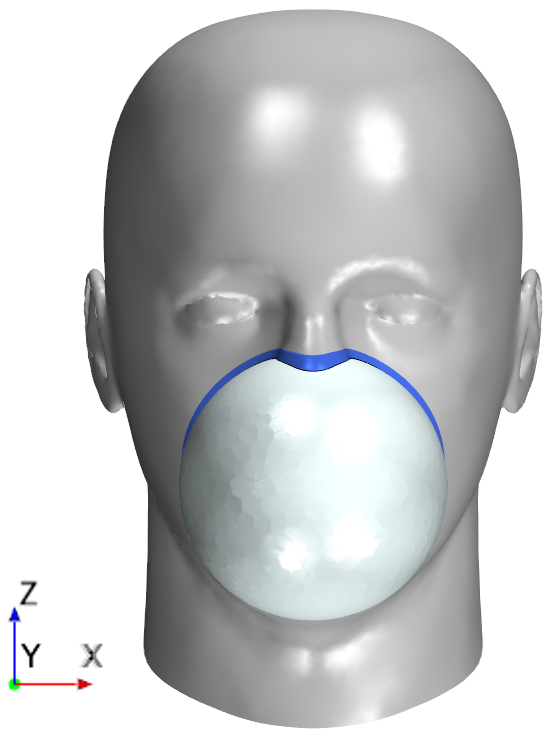} \\
\includegraphics[width=\linewidth]{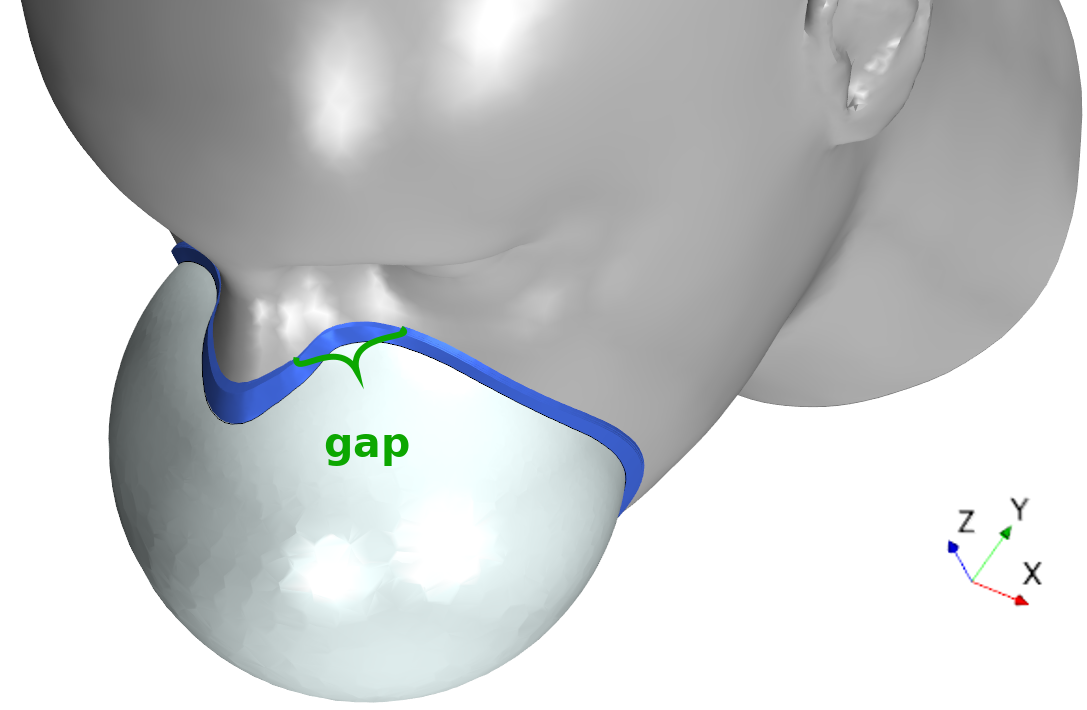} 
\caption{Top: Head (grey) with generic FFP3-type mask with filter surface (white) and seal (blue); bottom: located between nose and left eye, there is a gap with width $B_{\mathrm{g}}\approx 2.5\, \mathrm{cm}$ and height $H_{\mathrm{g}}$ between face and seal, through which air can leak} \label{FIG3Dgeom}
\end{center}
\end{figure}

The simulation setup was the same as in the 2D-flow simulations, with the following exceptions:  Head and mask had  realistic geometries as shown in Fig. \ref{FIG3Dgeom}, and the  air flows through the nostrils. The human head\footnote{Human Head 1/6 Scale by TheNewBlood, Creative Commons - Attribution - Non-Commercial license, Link: \url{http://www.thingiverse.com/thing:2859425}, license link: \url{http://creativecommons.org/licenses/by-nc/3.0/}} had a vertical distance between top of head and chin of $\approx 23\, \mathrm{cm}$.  The face mask had a spherical shape with radius $r=0.052\, \mathrm{m}$ and filter surface area $S_{\mathrm{m}}=0.02012\, \mathrm{m^{2}}$. The mask had a seal with length $L_{\mathrm{g}} \approx 1.2\, \mathrm{cm}$ and there was a gap between seal and face of average height $H_{\mathrm{g}} \approx 0.26\, \mathrm{mm}$, which extended  along a width of $B_{\mathrm{g}}\approx 2.5\, \mathrm{cm}$ below the left eye (cf. Fig. \ref{FIG3Dgeom}).

The solution domain was a box with dimensions $-0.246\, \mathrm{m} \leq x \leq 0.246\, \mathrm{m}$, $-0.35\, \mathrm{m} \leq y \leq 0.178\, \mathrm{m}$, $-0.146\, \mathrm{m} \leq z \leq 0.43\, \mathrm{m}$. The 3D-grid consisted of polyhedral cells with prism cells near the wall boundaries, so that there were more than $10$ cells per gap height.   The boundary opposite to the mask was the pressure outlet and all other boundaries except the mask surface and the nostrils were impermeable no-slip walls. The computational grid consisted of ca. $0.9$ million cells.

\begin{figure}[H]
\begin{center}
\includegraphics[width=0.9\linewidth]{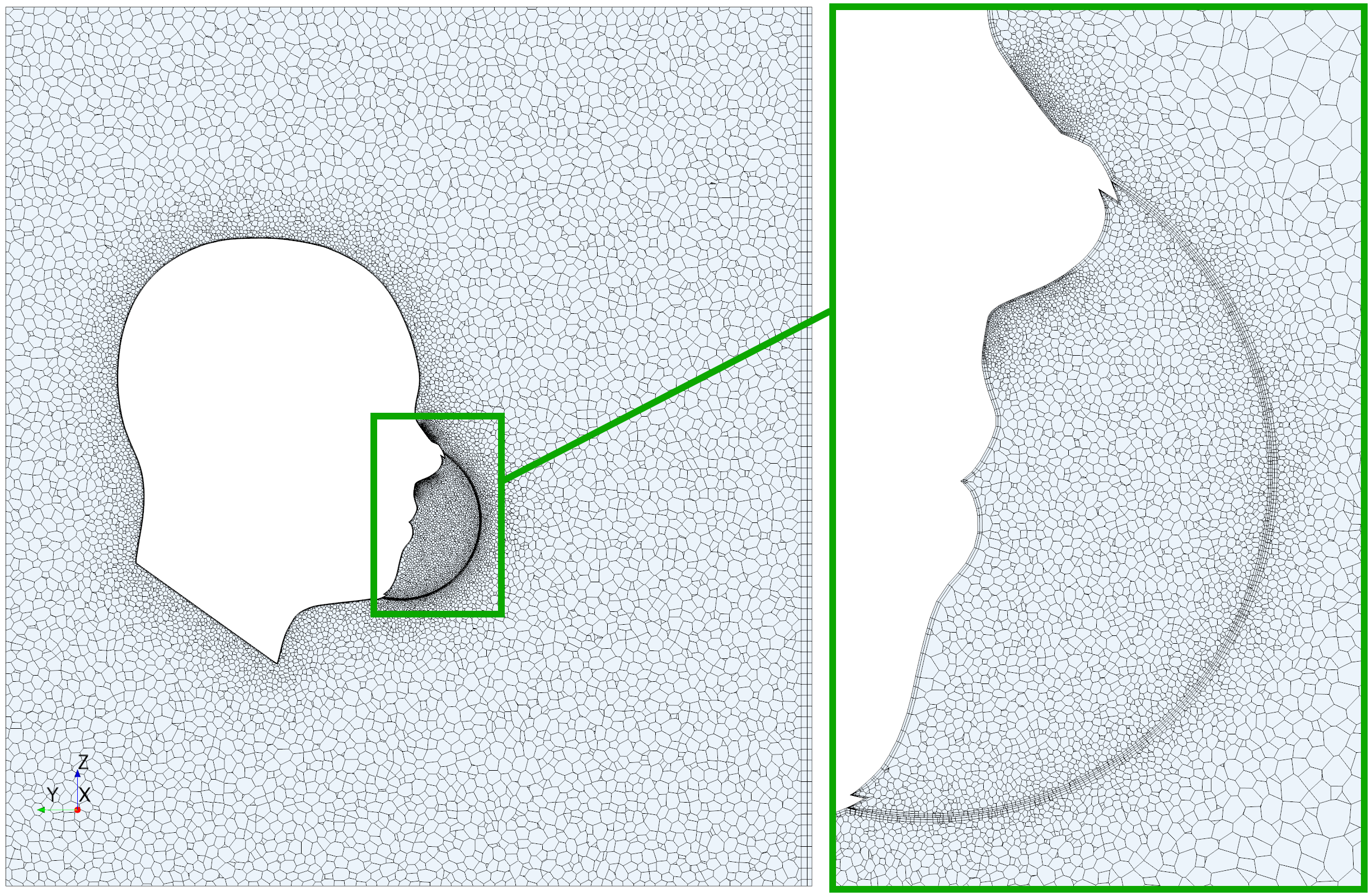}
\includegraphics[width=0.9\linewidth]{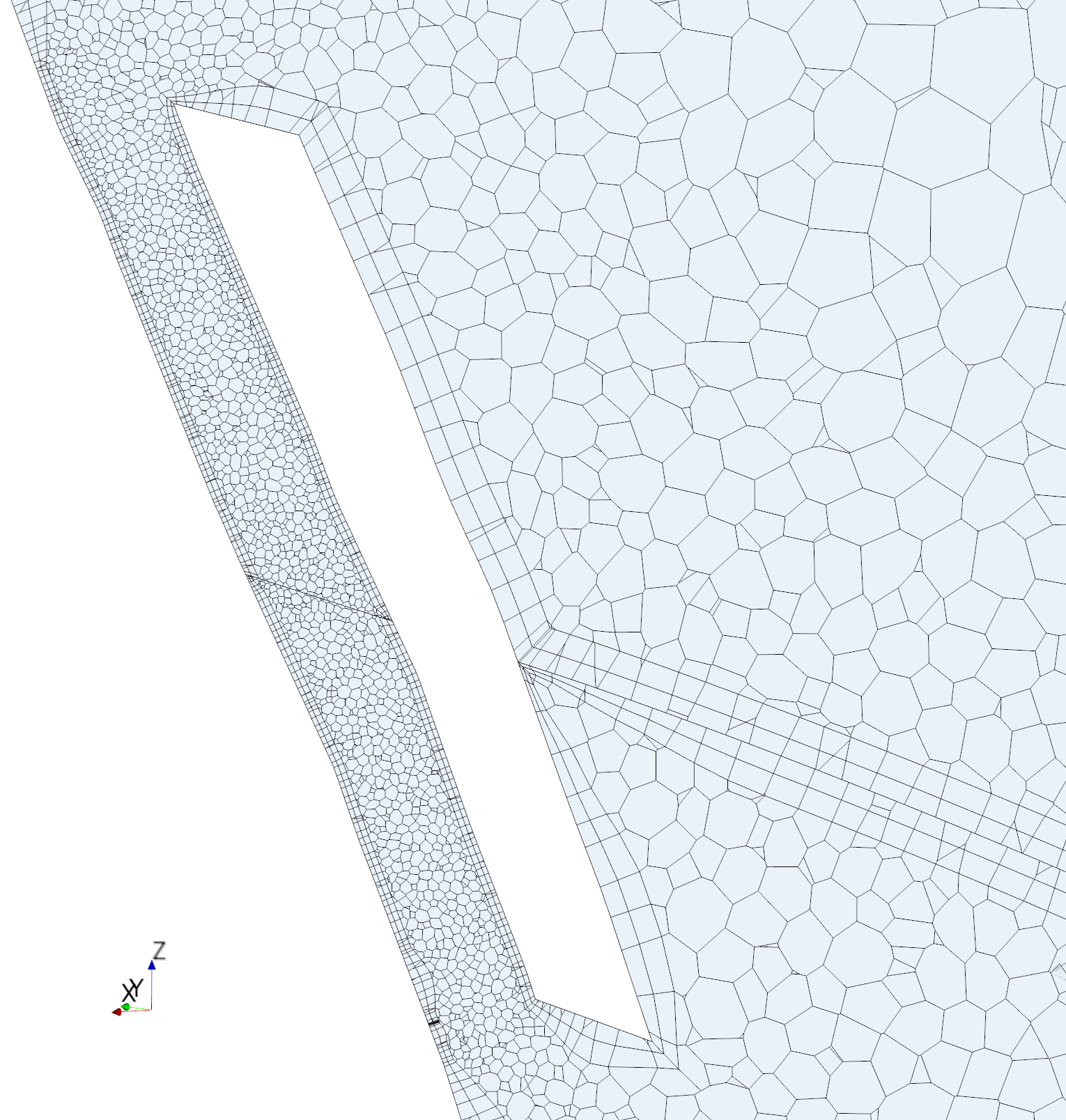}
\caption{Top: Computational grid for 3D-flow simulations with realistic head and mask geometries (top left) and close-up views of mask (top right) and gap (bottom)} \label{FIG3Dgrid2}
\end{center}
\end{figure}

Two further cases were investigated with a slightly modified setup, i.e. under-relaxation factors $0.9$ (velocities) and $0.5$ (pressure), time step $\Delta t = 0.0002\, \mathrm{s}$ and a finer grid with ca. $3\cdot 10^{6}$ cells (cf. Fig. \ref{FIG3Dgrid2}). 

In one set of simulations, air was allowed to enter the mask through two gaps below the eyes, each with width $B_{\mathrm{g}}=1.08\, \mathrm{cm}$ and average height $H_{\mathrm{g}}=0.66\, \mathrm{mm}$. The total gap length was $L_{\mathrm{g}}=0.71\, \mathrm{cm}$, which corresponds more closely to a typical mask seal. Both inhaling and exhaling were simulated with total flow rates $F_{\mathrm{t}}= 30\, \mathrm{L/min}$ and $F_{\mathrm{t}}= 95\, \mathrm{L/min}$.
%Simulations were also performed for a perfectly sealed mask.

In another set of simulations, below both eyes there was a gap with a circular-segment cross-section,  with total gap width $B_{\mathrm{g}}=2.32\, \mathrm{cm}$ and average gap height $ H_{\mathrm{g}} = 1.45\, \mathrm{mm}$ as shown in Fig. \ref{FIG3DexhaleIsoU}. Air was exhaled with total flow rates  $F_{\mathrm{t}}=30\, \mathrm{L/min}$ and $F_{\mathrm{t}}=95\, \mathrm{L/min}$.

\section{Results}
\label{SECres}
\begin{figure}[H]
\begin{center}
\begin{myfont}
\begin{small}
Flow rate $F_{\mathrm{t}}=30\, \mathrm{L/min}$\\
\includegraphics[width=\factor\linewidth]{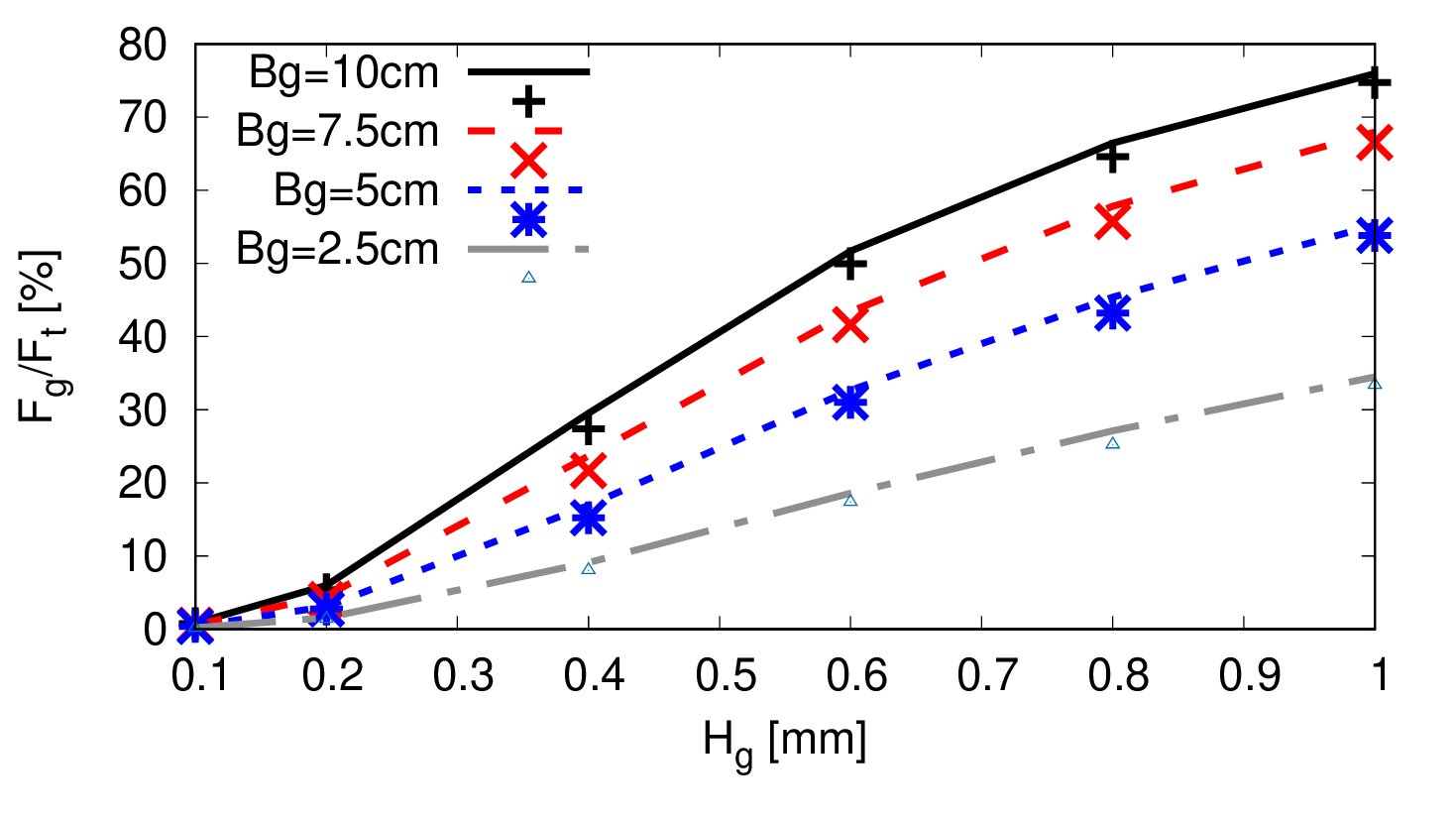} \\
Flow rate $F_{\mathrm{t}}=95\, \mathrm{L/min}$\\
\includegraphics[width=\factor\linewidth]{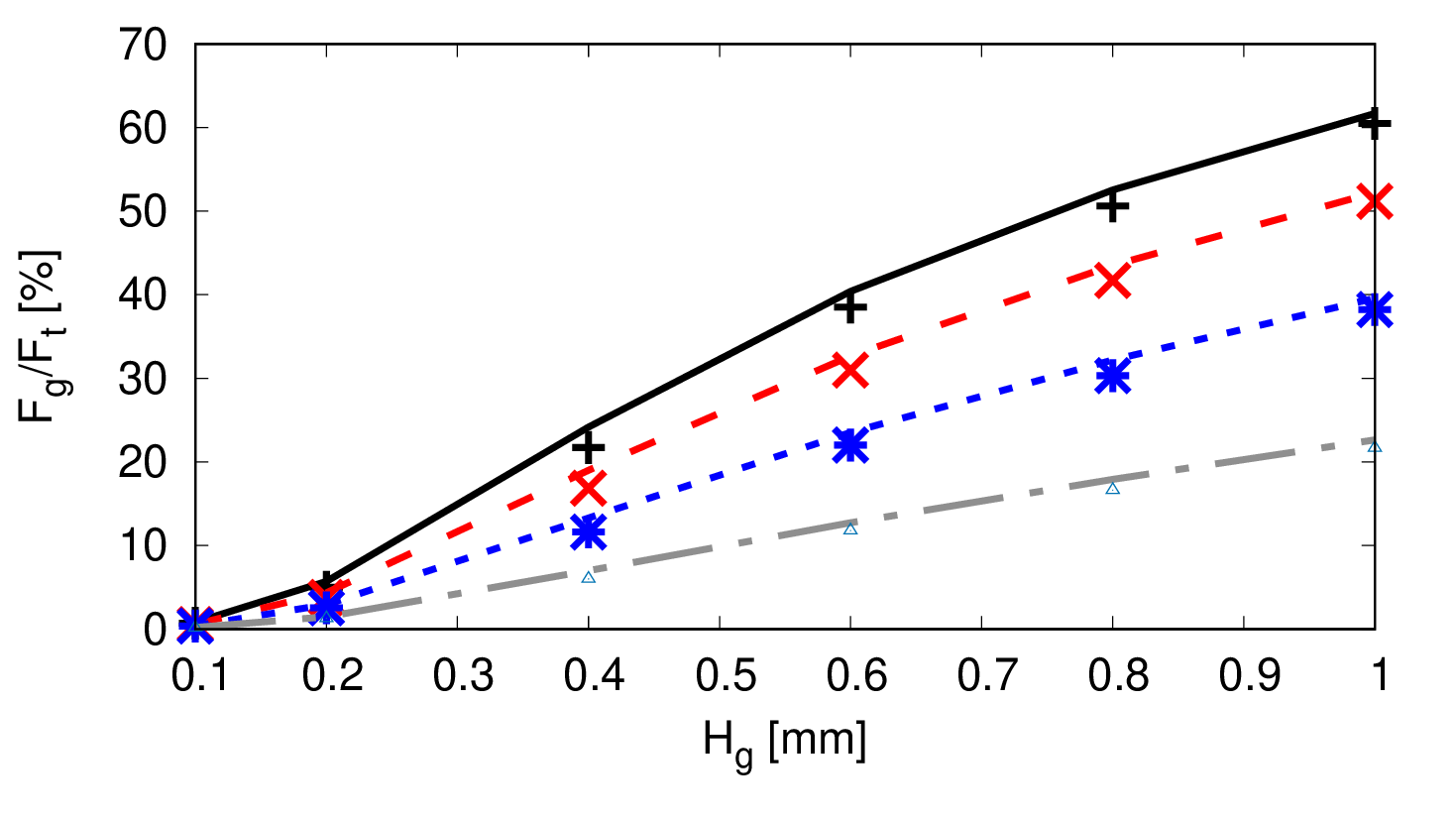} 
\end{small}
\end{myfont}
\caption{Analytical predictions and simulation results for the flow rate $F_{\mathrm{g}}$ through the gap  as percentage of the total flow rate $F_{\mathrm{t}}$, as a function of gap height $H_{\mathrm{g}}$, for a mask with viscous porous resistance $C_{\mathrm{m}}=2000\, \mathrm{m/s}$, gap length $L_{\mathrm{g}}=1\, \mathrm{cm}$ and different  gap widths $B_{\mathrm{g}}$; lines represent analytical predictions and points represent simulation results
} \label{FIGFt0Fg}
\end{center}
\end{figure}

\begin{figure}[H]
\begin{center}
\begin{myfont}
\begin{small}
Flow rate $F_{\mathrm{t}}=30\, \mathrm{L/min}$\\
\includegraphics[width=\factor\linewidth]{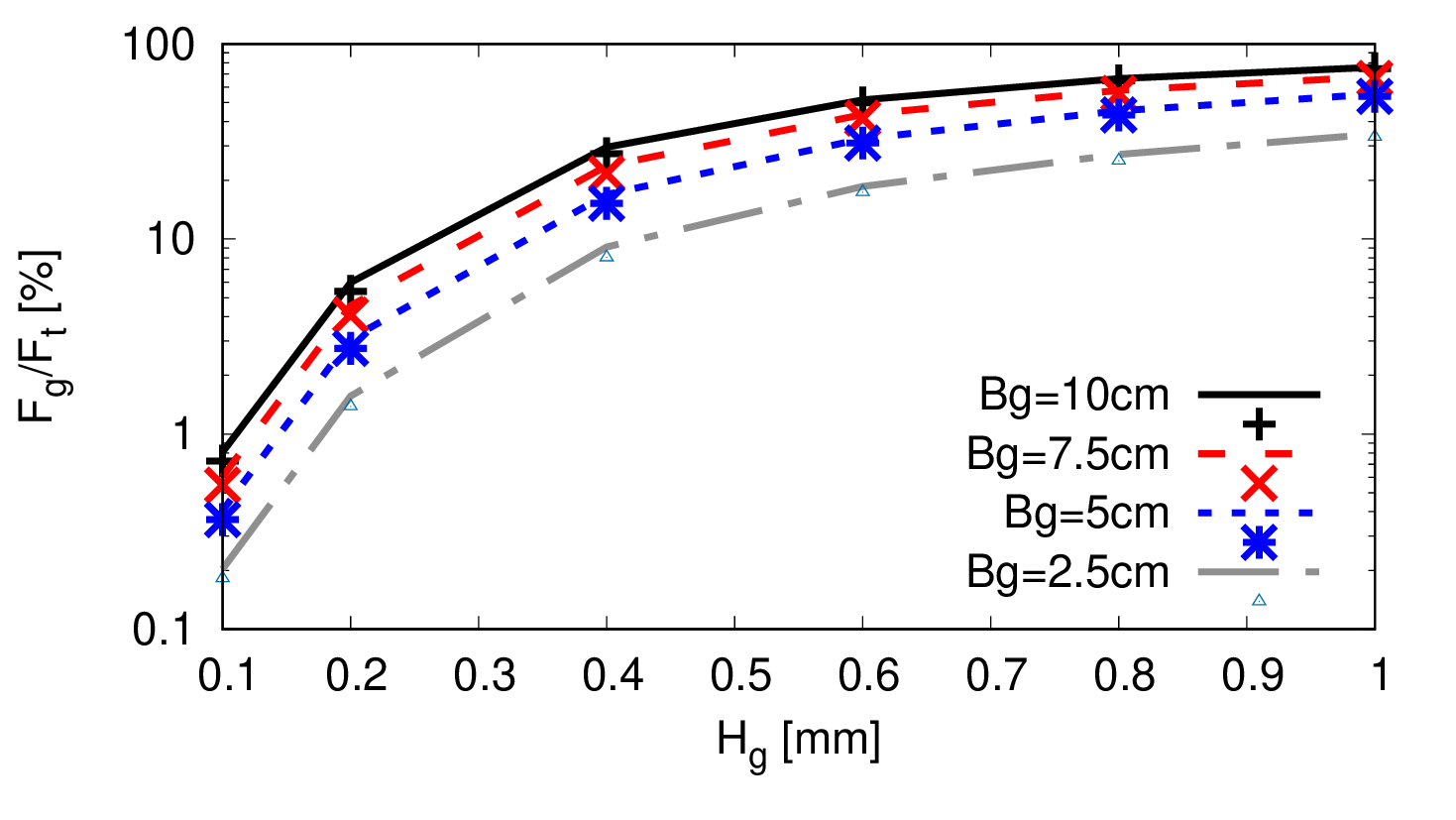} \\
Flow rate $F_{\mathrm{t}}=95\, \mathrm{L/min}$\\
\includegraphics[width=\factor\linewidth]{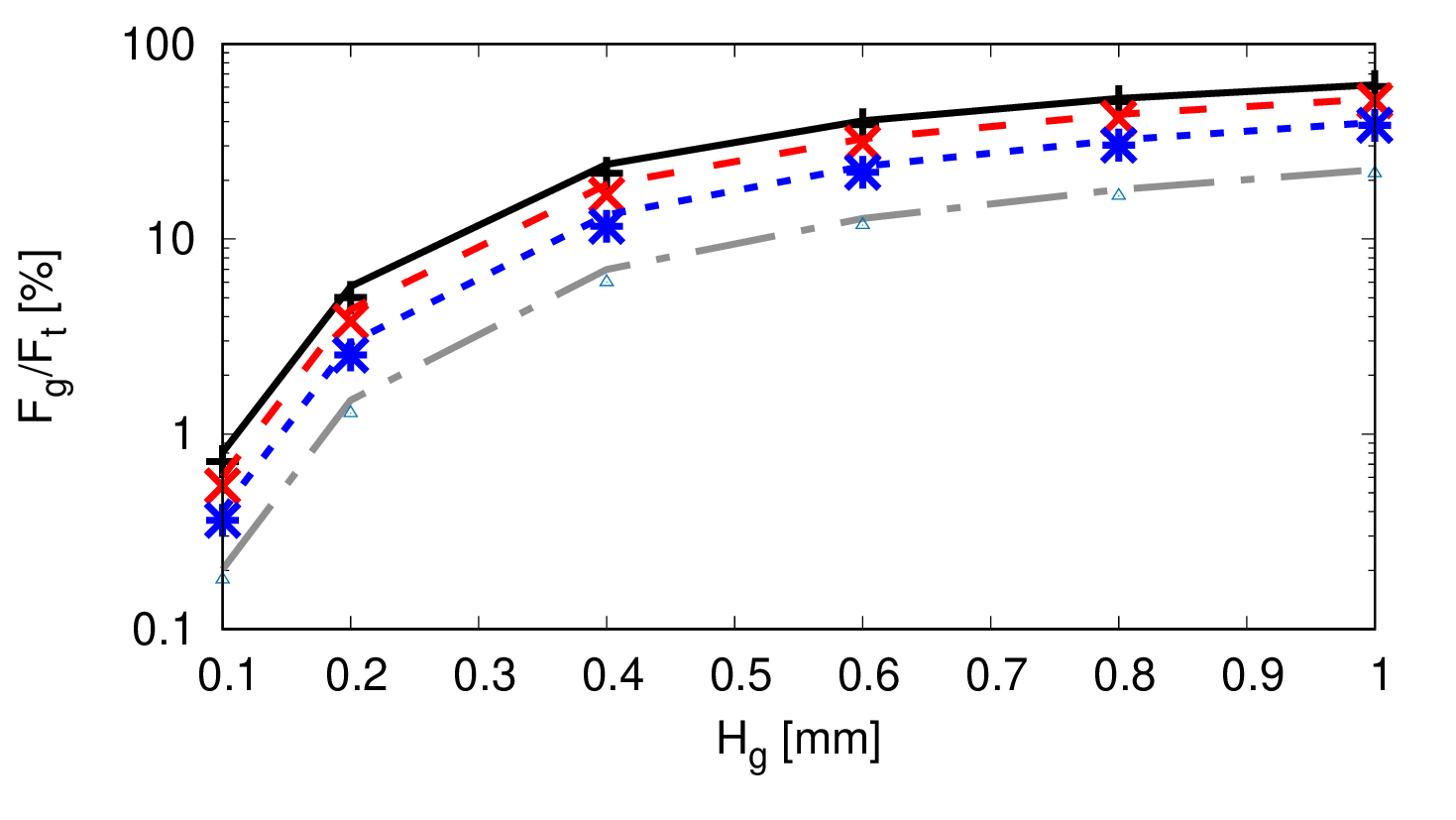} 
\end{small}
\end{myfont} 
\caption{As Fig. \ref{FIGFt0Fg}, except for logarithmic vertical axis} \label{FIGFt0FgLOG}
\end{center}
\end{figure}

\begin{figure}[H]
\begin{center}
\begin{myfont}
\begin{small}
Flow rate $F_{\mathrm{t}}=30\, \mathrm{L/min}$\\
\includegraphics[width=\factor\linewidth]{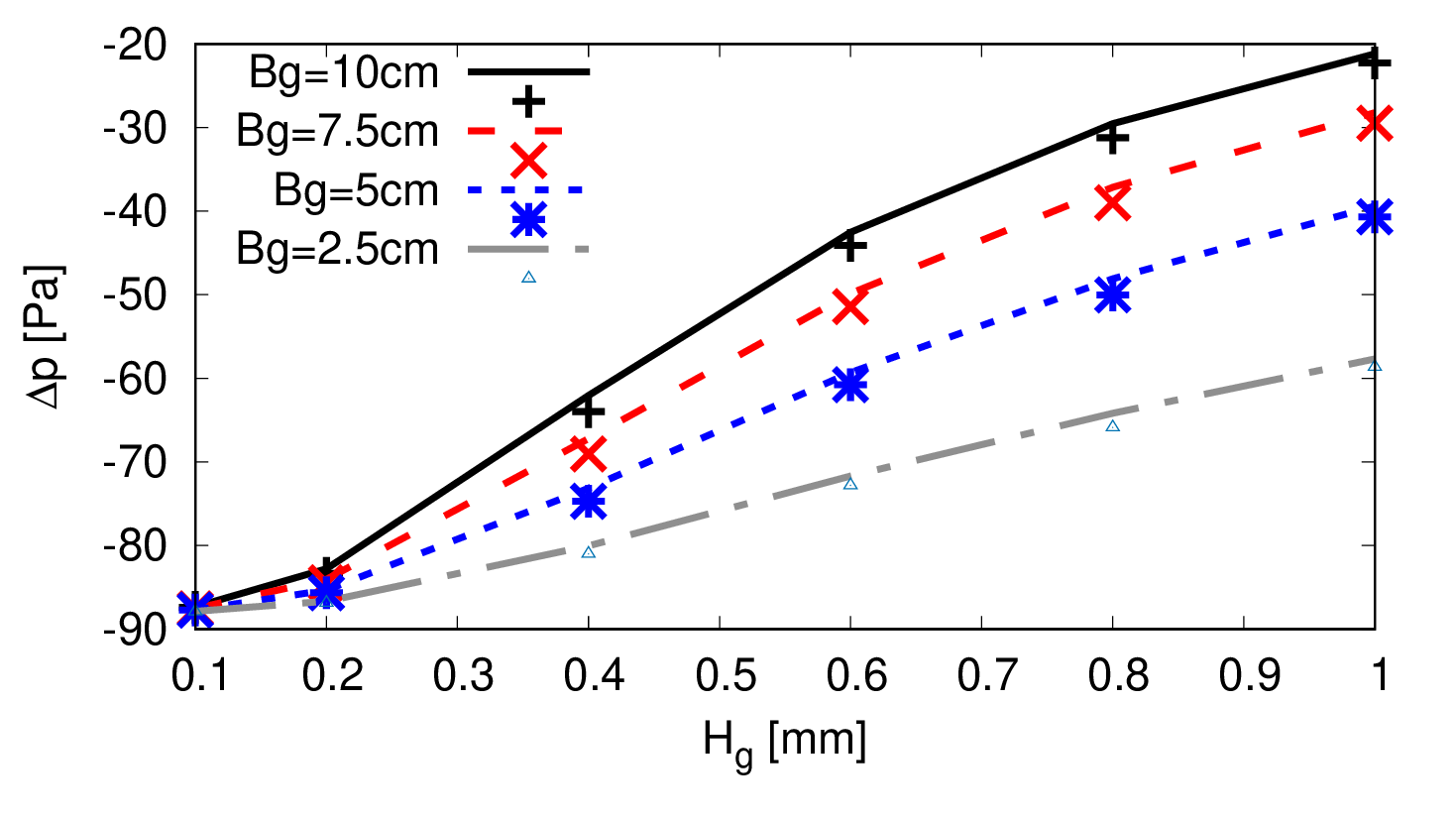} \\
Flow rate $F_{\mathrm{t}}=95\, \mathrm{L/min}$\\
\includegraphics[width=\factor\linewidth]{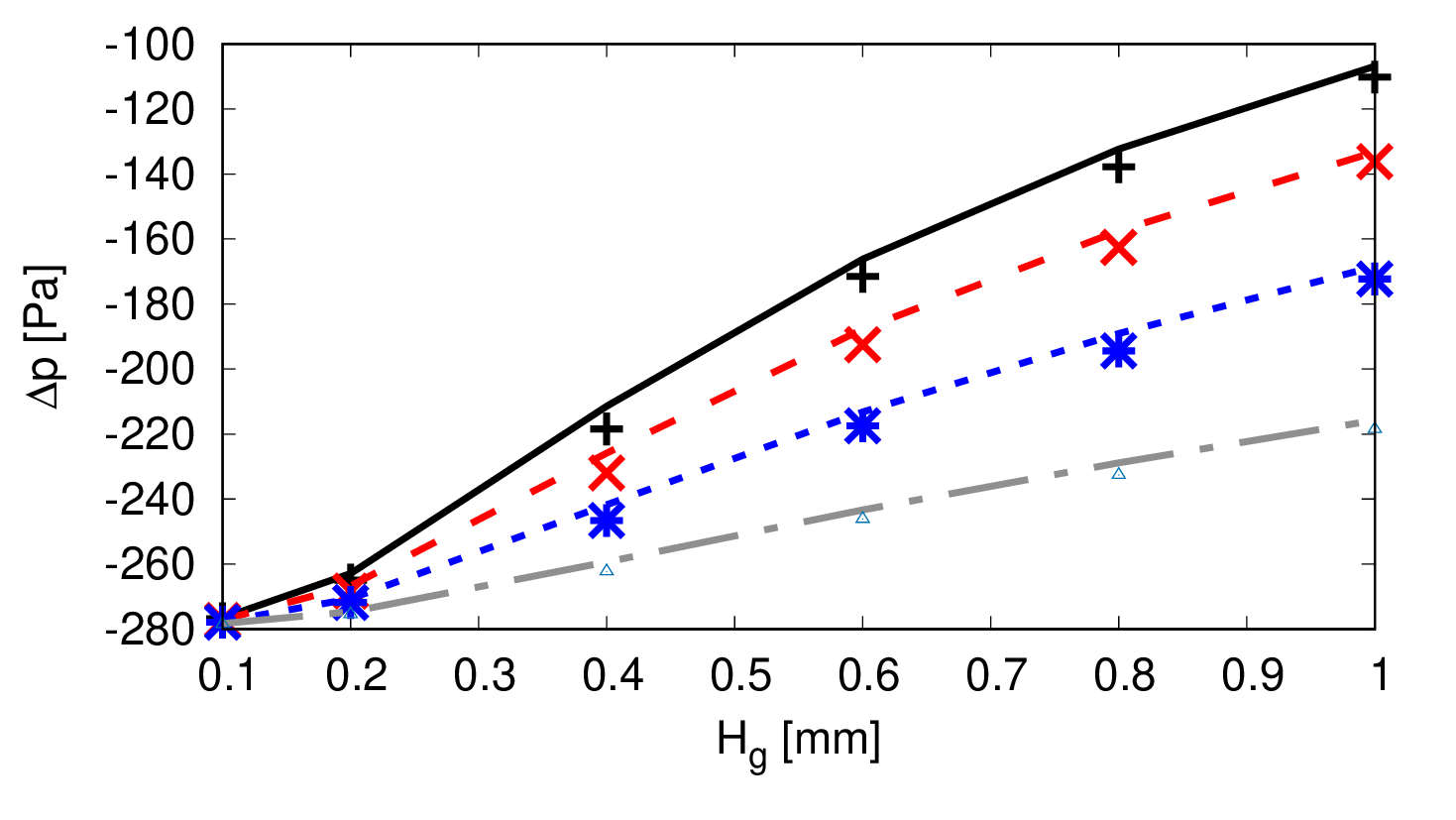} 
\end{small}
\end{myfont}
\caption{Analytical predictions and simulation results for pressure drop $\Delta p$  as a function of gap height $H_{\mathrm{g}}$, for different gap widths $B_{\mathrm{g}}$; lines represent analytical predictions and points represent simulation results} \label{FIGdp}
\end{center}
\end{figure}

Figures \ref{FIGFt0Fg}-\ref{FIGdp} show results from 2D-flow simulations based on the setup from Sect. \ref{SECsetup} for  inhaling. The mask has a viscous porous resistance of $C_{\mathrm{m}}=2000\, \mathrm{m/s}$, corresponding to a mask that, when tightly fitted, produces a pressure drop $\Delta p$ close to the upper limit for a FFP3 mask according to the EN149 norm, i.e. $\Delta p < 100\, \mathrm{Pa}$  (for total flow rate $F_{\mathrm{t}}=30\, \mathrm{L/min}$) and $\Delta p < 300\, \mathrm{Pa}$  (for $F_{\mathrm{t}}=95\, \mathrm{L/min}$). Typical pressure drops measured in experiments for FFP2- and FFP3-masks were e.g. $97\, \mathrm{Pa} < \Delta p < 244\, \mathrm{Pa}$ (Jung et al., 2014) and $150\, \mathrm{Pa} < \Delta p < 230\, \mathrm{Pa}$ (Serfozo et al., 2017), indicating that the choice of $C_{\mathrm{m}}$ is feasible.

Figure \ref{FIGFt0Fg} shows that the flow rate through the gap between mask seal and face depends non-linearly on the gap height $H_{\mathrm{g}}$. The predictions from the simple analytical model from Sect. \ref{SECtheory} agree well with the more sophisticated 2D-flow simulation results.

Figure \ref{FIGFt0FgLOG} shows that gap heights of $H_{\mathrm{g}} \approx 0.1\, \mathrm{mm}$ or less are required so that less than $1\%$ of the air flows through the gap. Figure \ref{FIGFt0FgLOG} also shows that even a gap of height $H_{\mathrm{g}} = 0.2\, \mathrm{mm}$ (which corresponds to the height of a beard  a few hours after a close shave) causes that $2\%$ to $8\%$ of the inhaled air flows unfiltered through the gap, depending on the width of the gap (here: between $2.5\, \mathrm{cm}$ and $10\, \mathrm{cm}$). 

The results also show that the gap height $H_{\mathrm{g}}$ has the largest influence: increasing gap height $H_{g}$ by a factor of $2$ can increase the flow rate $F_{\mathrm{g}}$ through the gap by a factor of up to $10$. The gap width had a comparatively small influence: increasing gap width $B_{g}$ by a factor of $2$ increased the flow rate $F_{\mathrm{g}}$ through the gap by a factor of up to ca. $2$.

\begin{figure}[H]
\begin{center}
\includegraphics[width=0.8\linewidth]{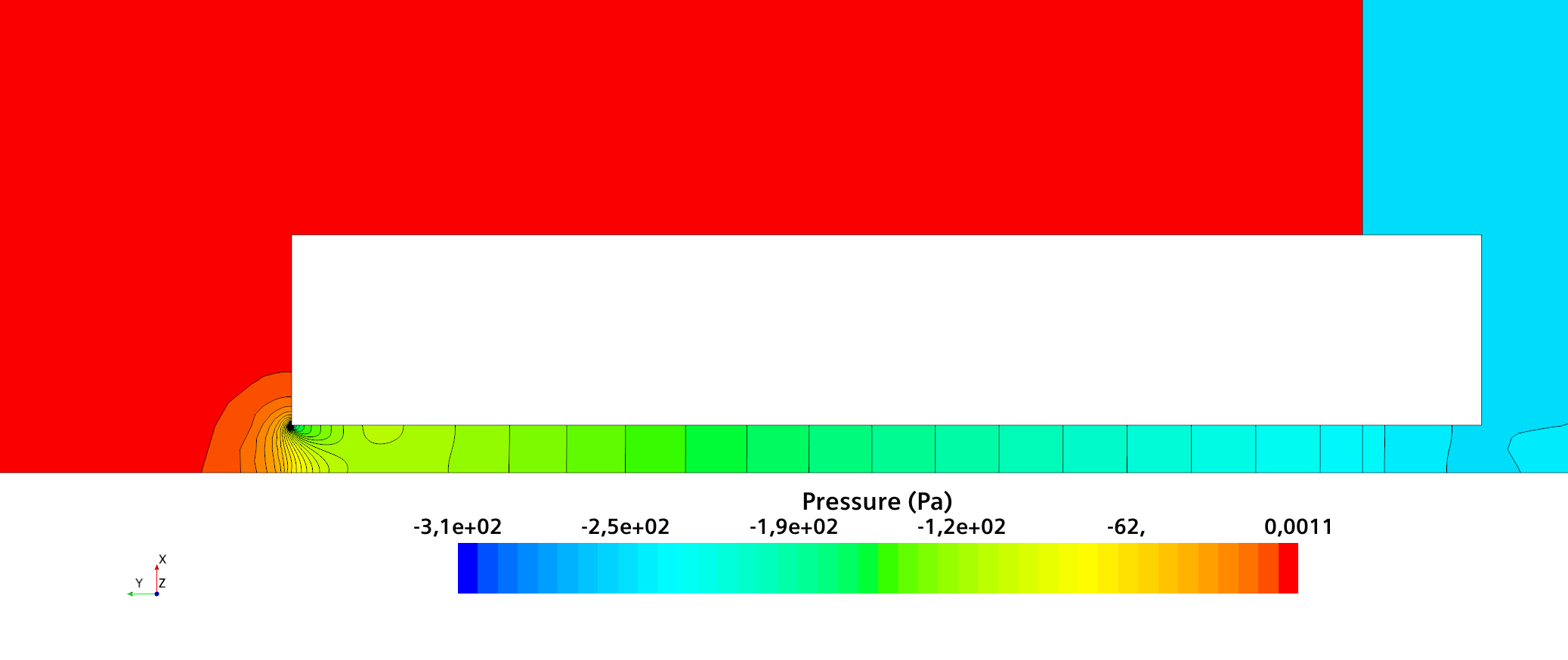} \\
\includegraphics[width=0.8\linewidth]{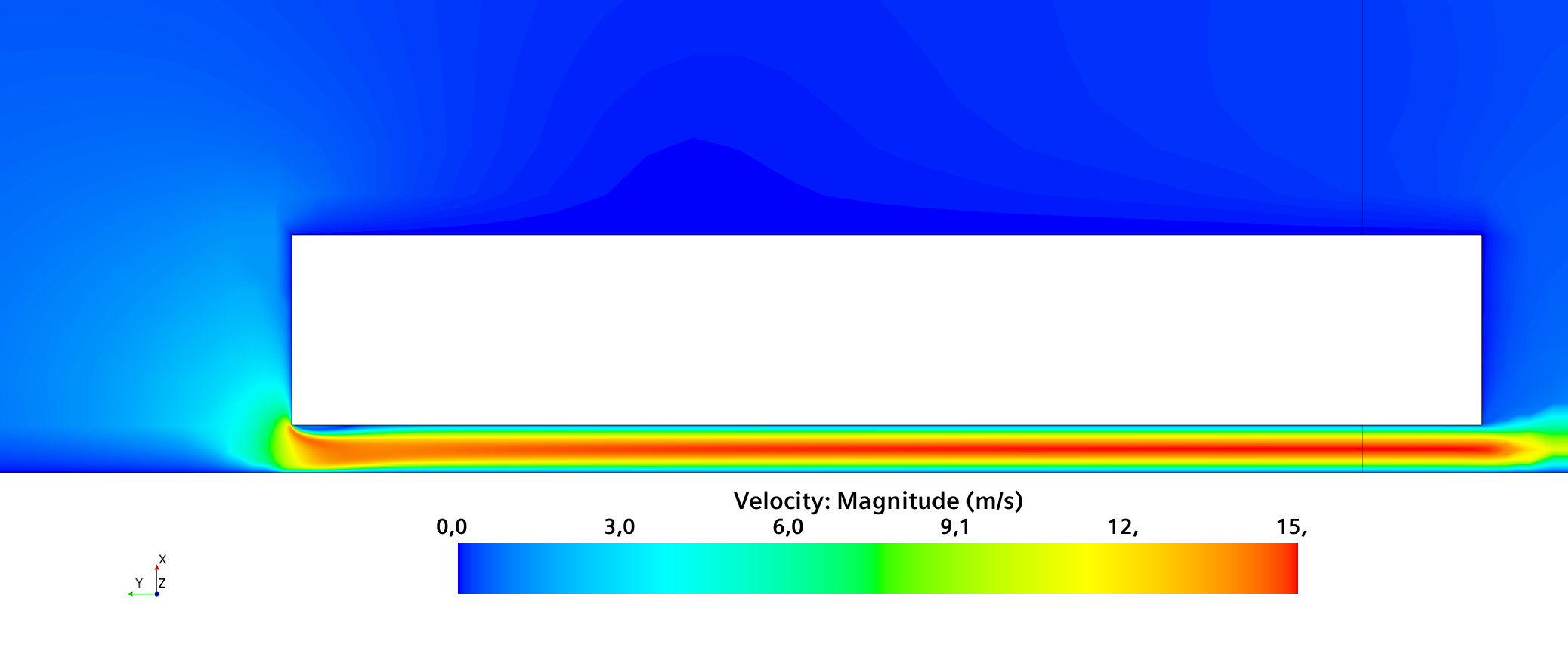} \\
\includegraphics[width=0.8\linewidth]{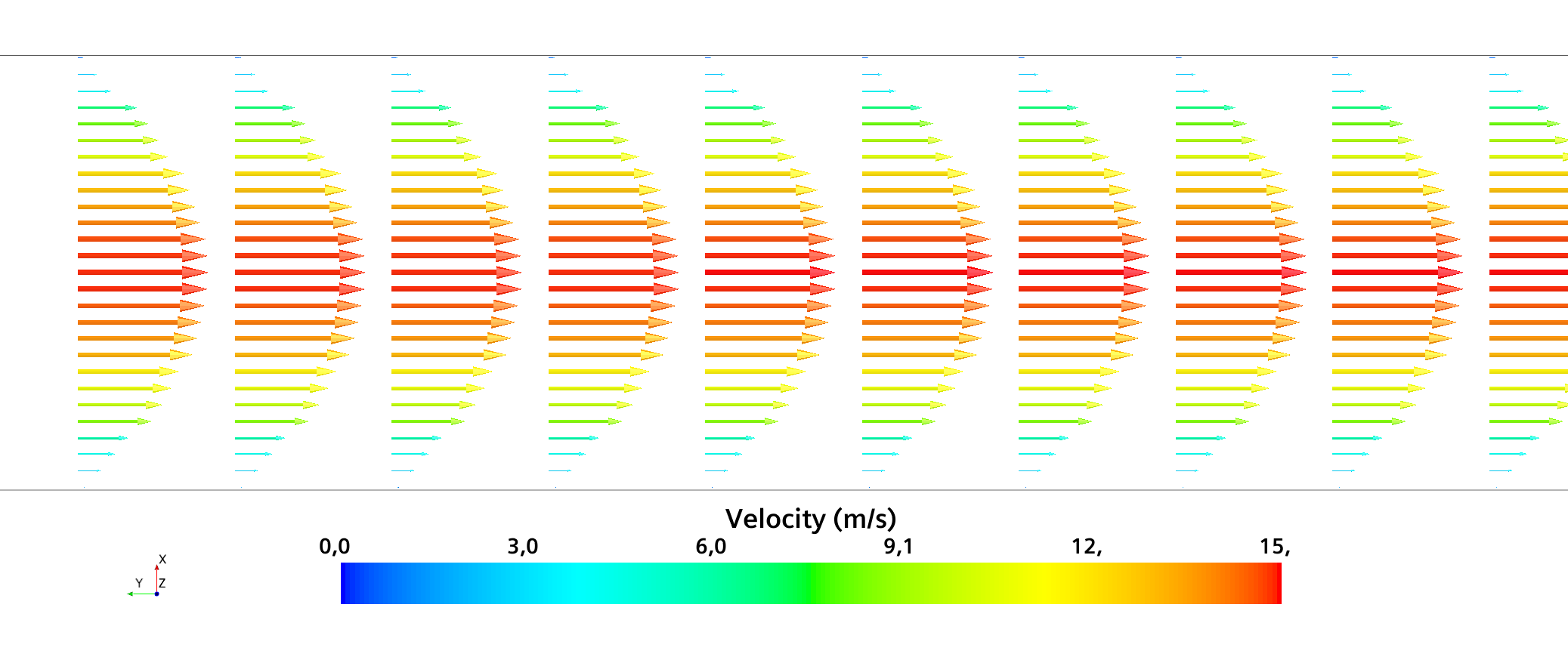} 
\caption{Top: Pressure contours within the gap; middle: velocity magnitude inside the gap; bottom: close-up of center-part of the gap with velocity vectors; for simulation from Fig. \ref{FIGFt0Fg} with flow rate $F_{\mathrm{t}}=95\, \mathrm{L/min}$,  viscous porous filter resistance $C_{\mathrm{m}}=2000\, \mathrm{m/s}$, gap height $H_{\mathrm{g}}=0.4\, \mathrm{mm}$, gap length $L_{\mathrm{g}}=1\, \mathrm{cm}$ and gap width $B_{\mathrm{g}}=5\, \mathrm{cm}$} \label{FIG2DpvelVecs}
\end{center}
\end{figure}

Figure \ref{FIGdp} shows that the pressure drop across the mask filter can decrease by more than $75\%$ when the gap height is increased to $H_{\mathrm{g}}=1\, \mathrm{mm}$. This makes breathing easier, which could tempt mask wearers not to tighten the mask to the face, with the consequence that up to ca. $70\%$ of the air flows unfiltered through the gap into the mask. Figure \ref{FIG2DpvelVecs} shows pressure and velocities within the gap for a representative configuration.

\begin{figure}[H]
\begin{center}
\begin{myfont}
\begin{small}
Flow rate $F_{\mathrm{t}}=30\, \mathrm{L/min}$\\
\includegraphics[width=\factor\linewidth]{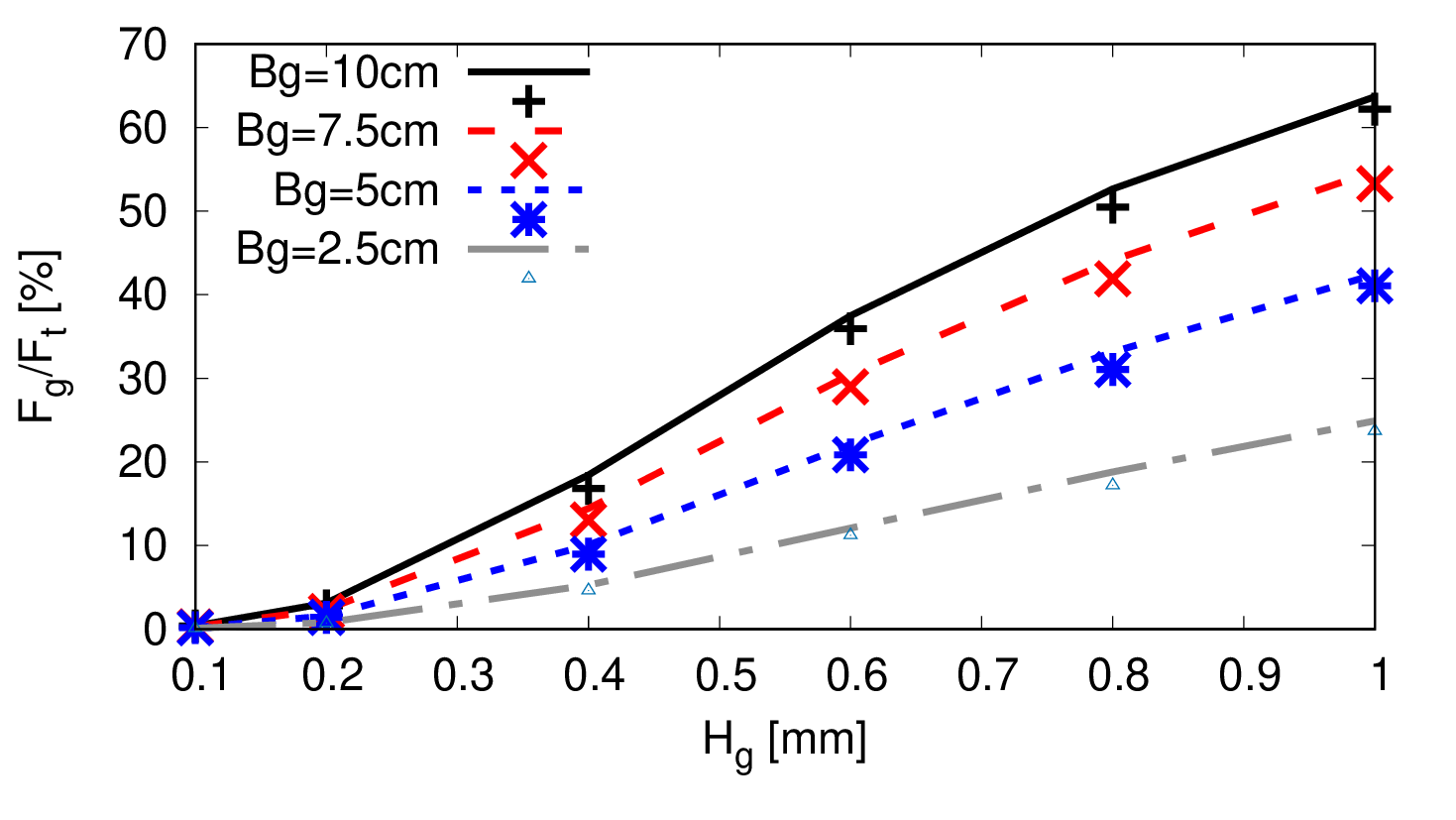} \\
Flow rate $F_{\mathrm{t}}=95\, \mathrm{L/min}$\\
\includegraphics[width=\factor\linewidth]{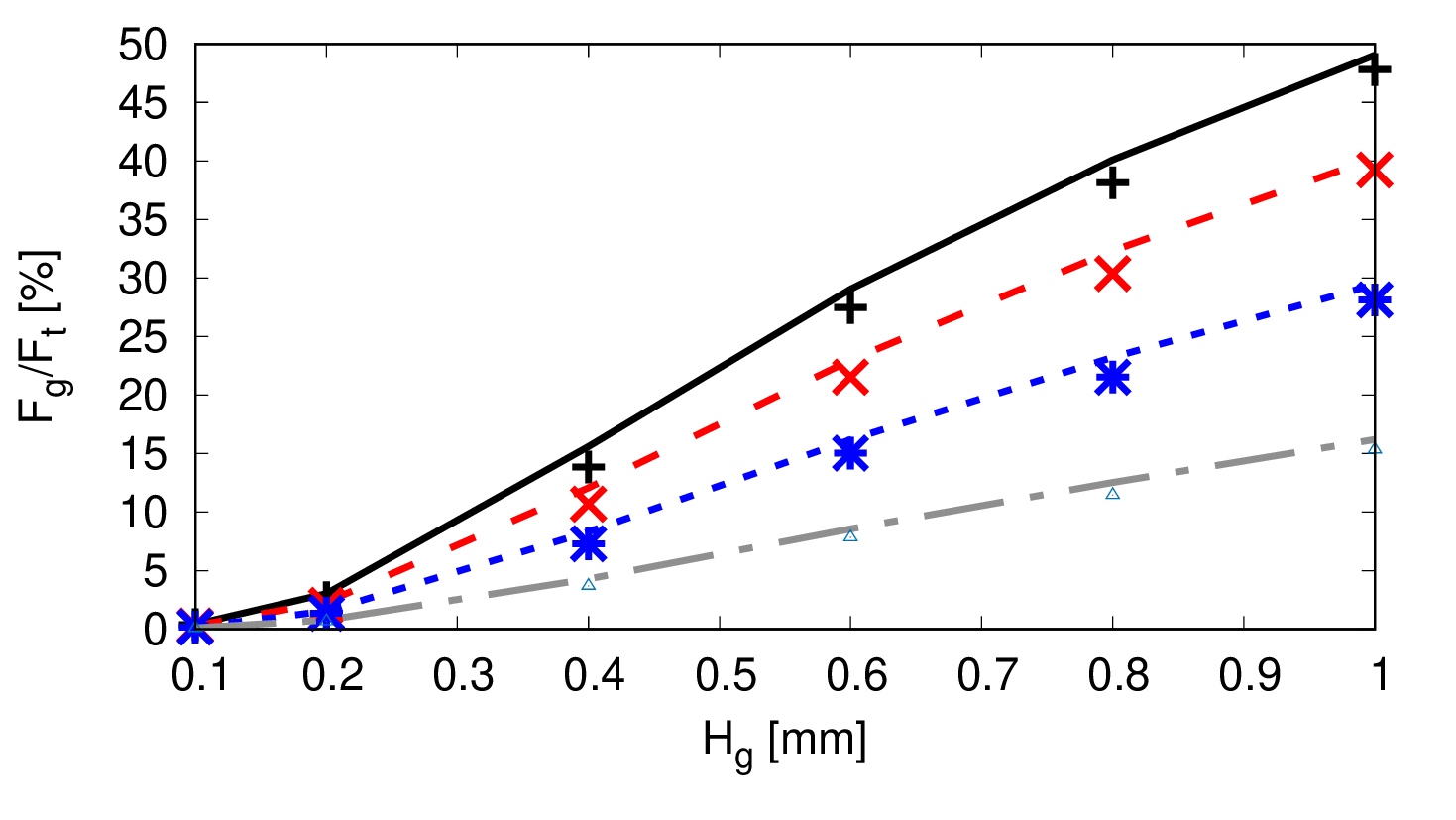} 
\end{small}
\end{myfont}
\caption{As Fig. \ref{FIGFt0Fg}, except for a mask with viscous porous resistance $C_{\mathrm{m}}=1000\, \mathrm{m/s}$} \label{FIGFt0FgCm1000}
\end{center}
\end{figure}

\begin{figure}[H]
\begin{center}
\begin{myfont}
\begin{small}
Flow rate $F_{\mathrm{t}}=30\, \mathrm{L/min}$\\
\includegraphics[width=\factor\linewidth]{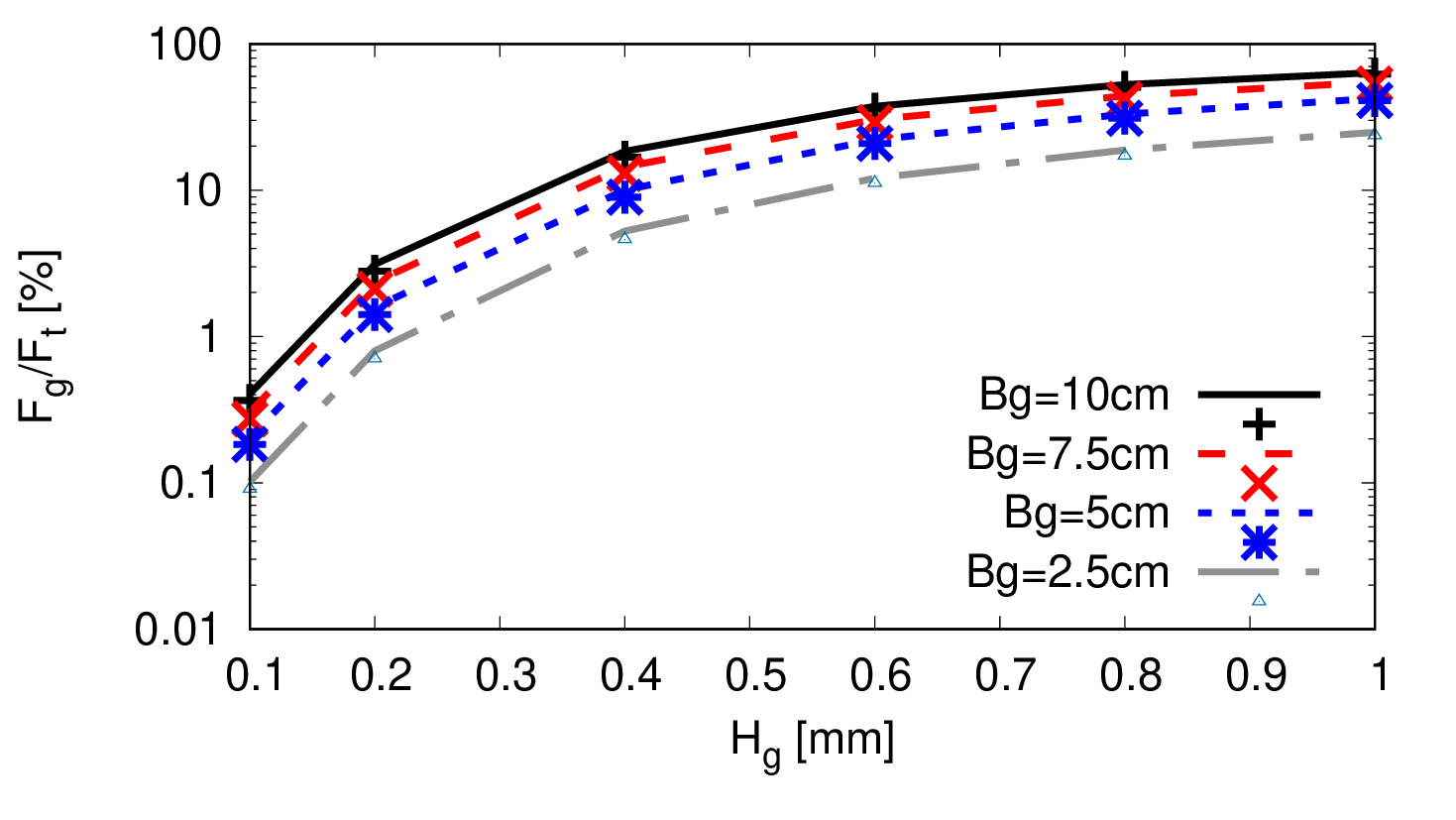} \\
Flow rate $F_{\mathrm{t}}=95\, \mathrm{L/min}$\\
\includegraphics[width=\factor\linewidth]{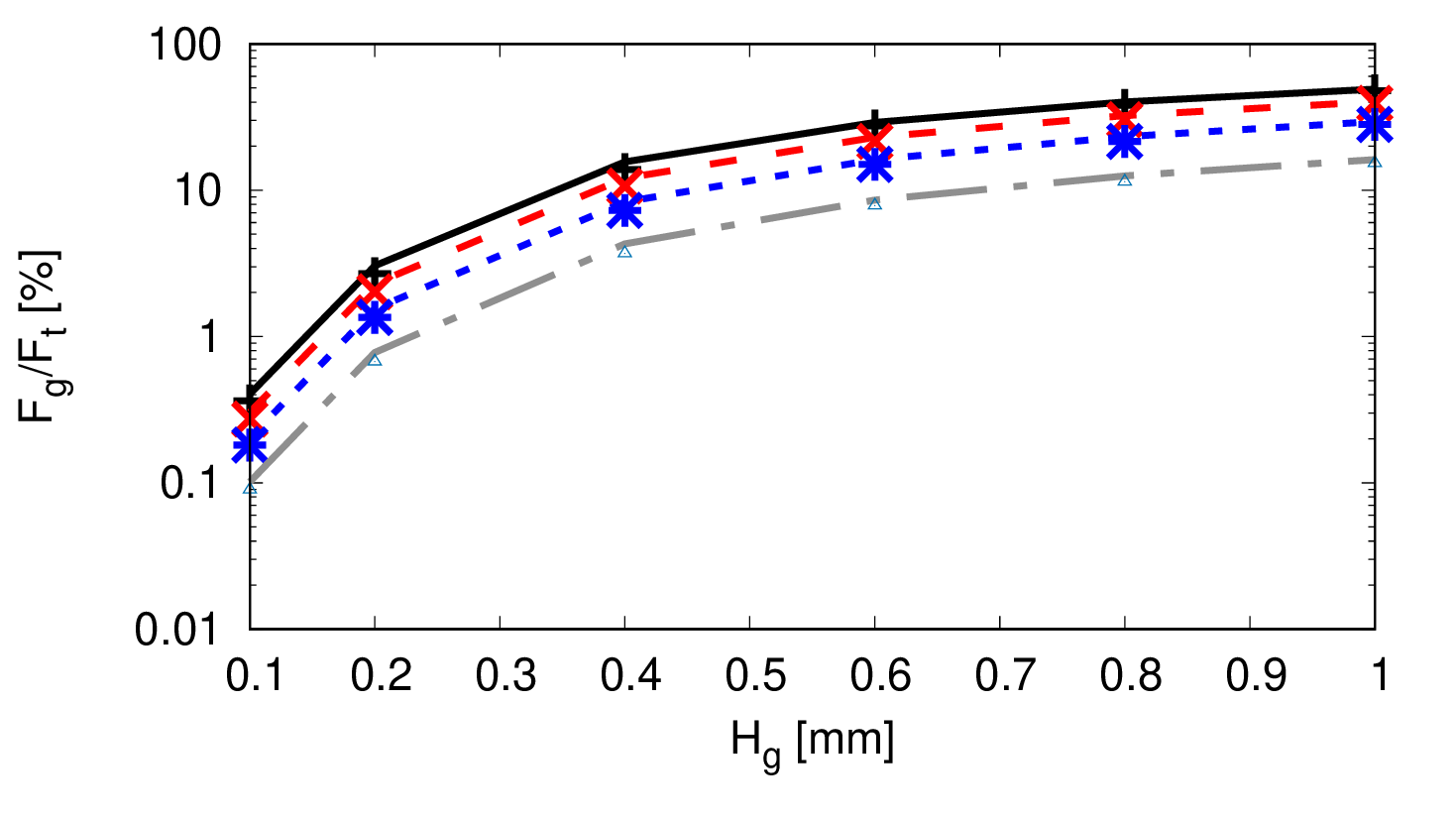} 
\end{small}
\end{myfont}
\caption{As Fig. \ref{FIGFt0FgCm1000}, except for logarithmic vertical axis} \label{FIGFt0FgLOGCm1000}
\end{center}
\end{figure}

\begin{figure}[H]
\begin{center}
\begin{myfont}
\begin{small}
Flow rate $F_{\mathrm{t}}=30\, \mathrm{L/min}$\\
\includegraphics[width=\factor\linewidth]{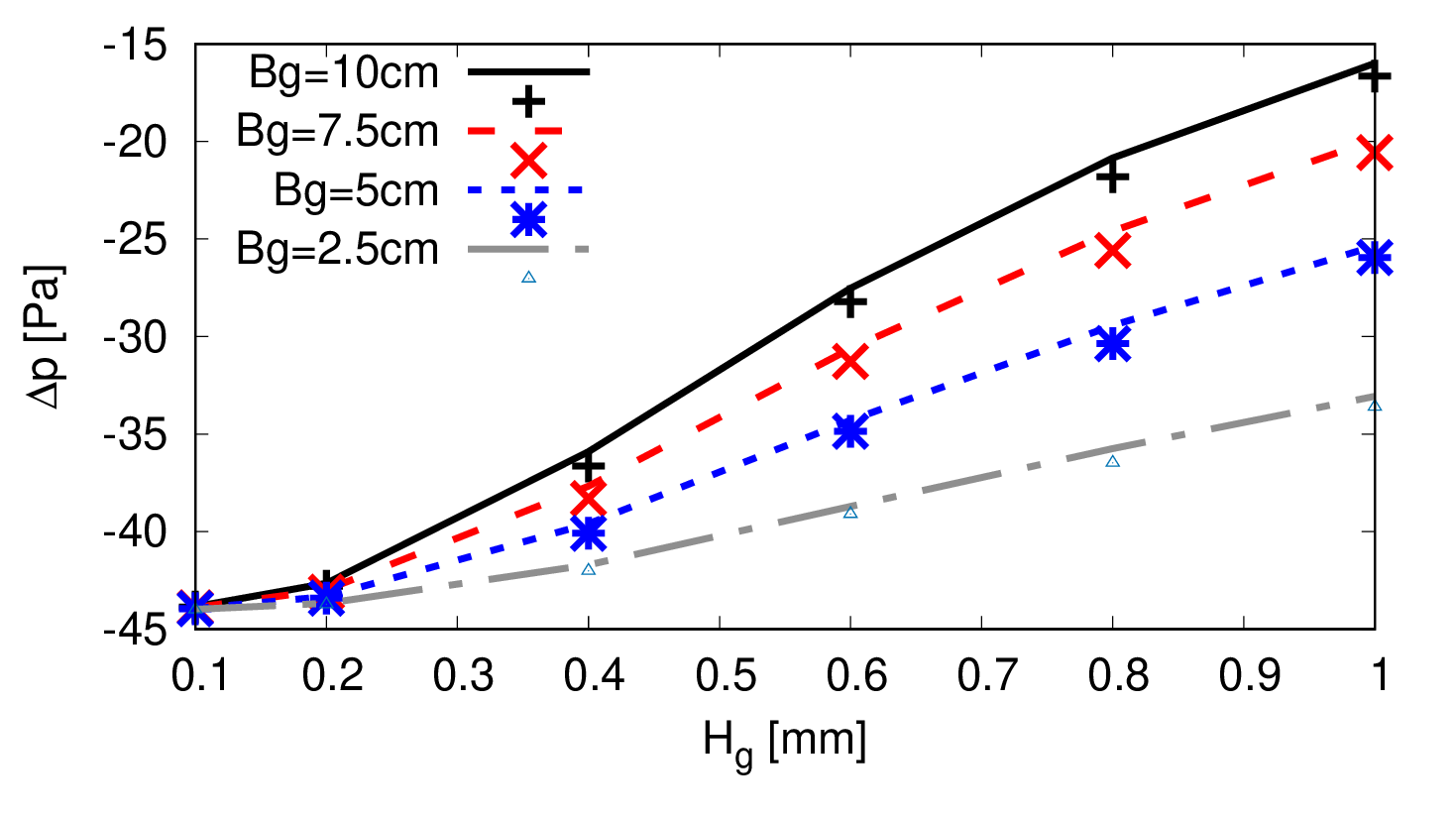} \\
Flow rate $F_{\mathrm{t}}=95\, \mathrm{L/min}$\\
\includegraphics[width=\factor\linewidth]{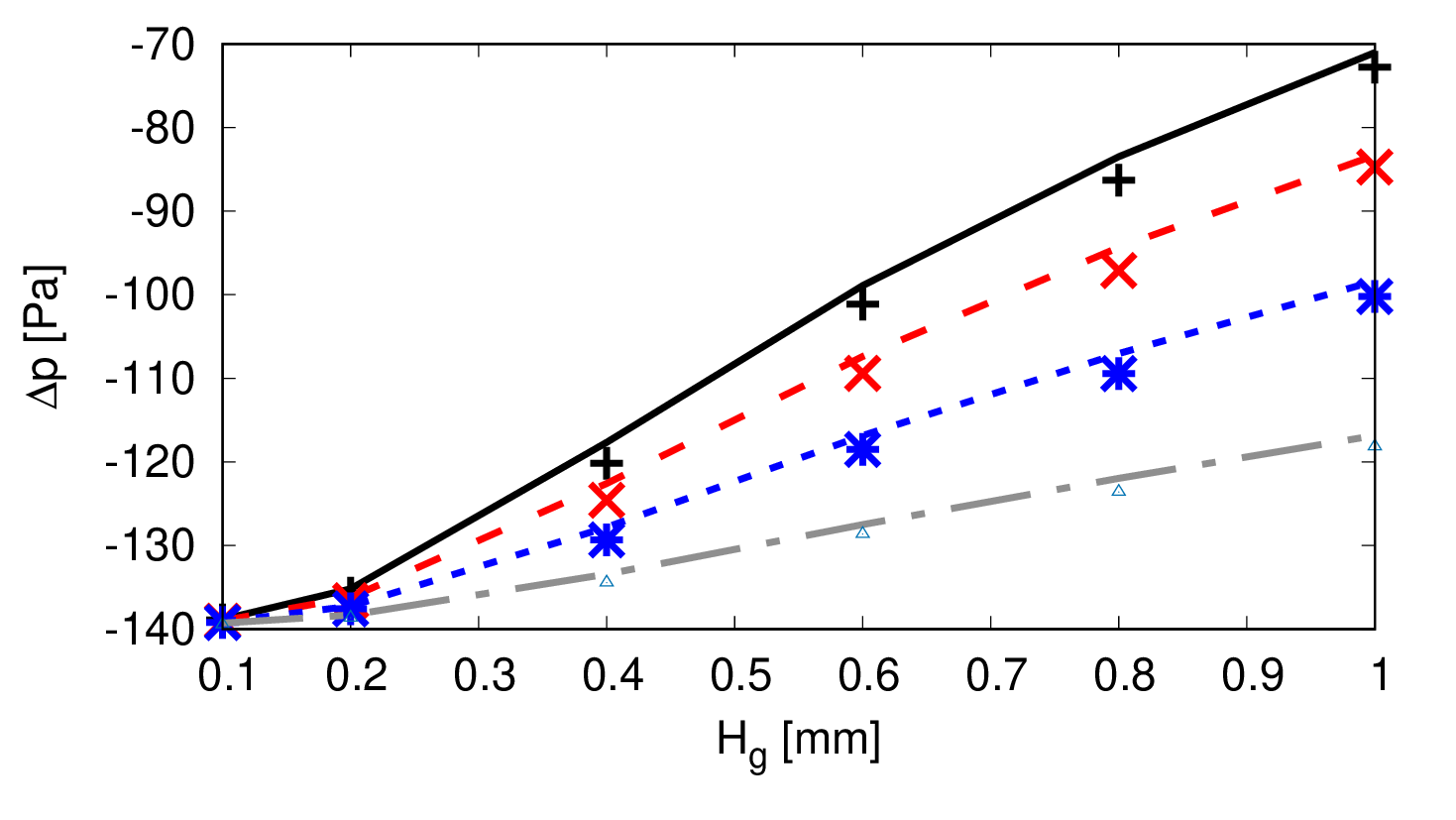} 
\end{small}
\end{myfont}
\caption{As Fig. \ref{FIGdp}, except for a mask with viscous porous resistance $C_{\mathrm{m}}=1000\, \mathrm{m/s}$} \label{FIGdpCm1000}
\end{center}
\end{figure}

Figures  \ref{FIGFt0FgCm1000}-\ref{FIGdpCm1000} show results for a mask with viscous porous resistance $C_{\mathrm{m}}=1000\, \mathrm{m/s}$. When tightly fitted, the mask produces a pressure drop close to the lowest values that FFP3-masks on the market may provide.  

The results show similar trends as those for $C_{\mathrm{m}}=2000\, \mathrm{m/s}$, and although the filter resistance is halved, the flow rate through the gap reduces only by $10-20\%$ in most cases.
Thus, the choice of filter material has a comparatively small influence concerning how large the gap height $H_{\mathrm{g}}$ may be before the mask ceases to fulfill  FFP3 requirements. 

Figures \ref{FIGFt0Fg}-\ref{FIGdpCm1000} show the results obtained for the fine grid. Results for medium and coarse grids are not depicted because they show no noticeable differences.
Average differences between simulation results for the flow rates on coarse, medium and fine grids are less than $0.8\%$. The differences between analytical predictions and simulation results for the flow rates was $<3\%$ for all simulated cases.

\begin{figure}[H]
\begin{center}
\begin{myfont}
\begin{small}
Gap width $B_{\mathrm{g}}=2\, \mathrm{cm}$\\
\includegraphics[width=\factor\linewidth]{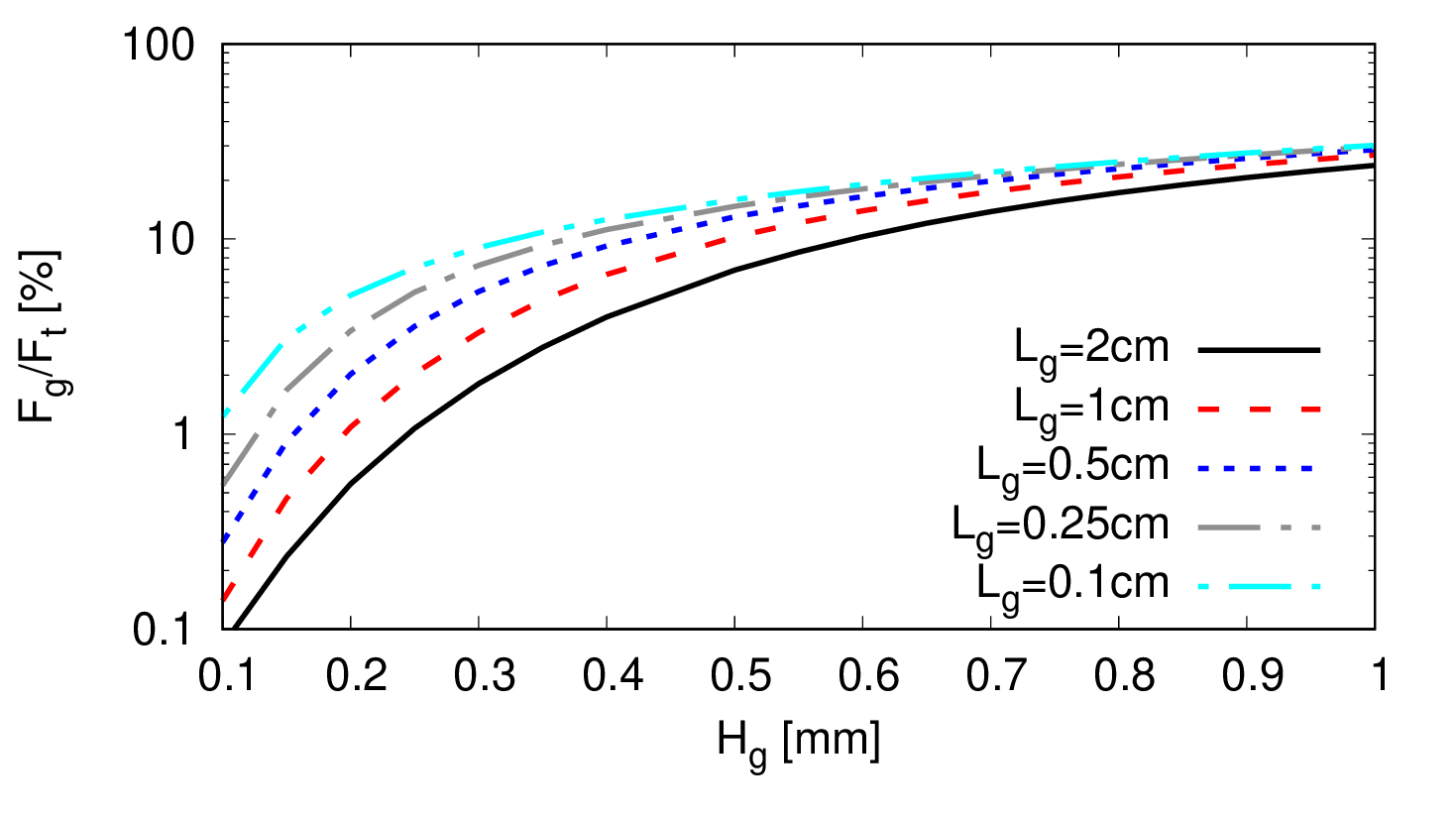} \\
Gap width $B_{\mathrm{g}}=10\, \mathrm{cm}$\\
\includegraphics[width=\factor\linewidth]{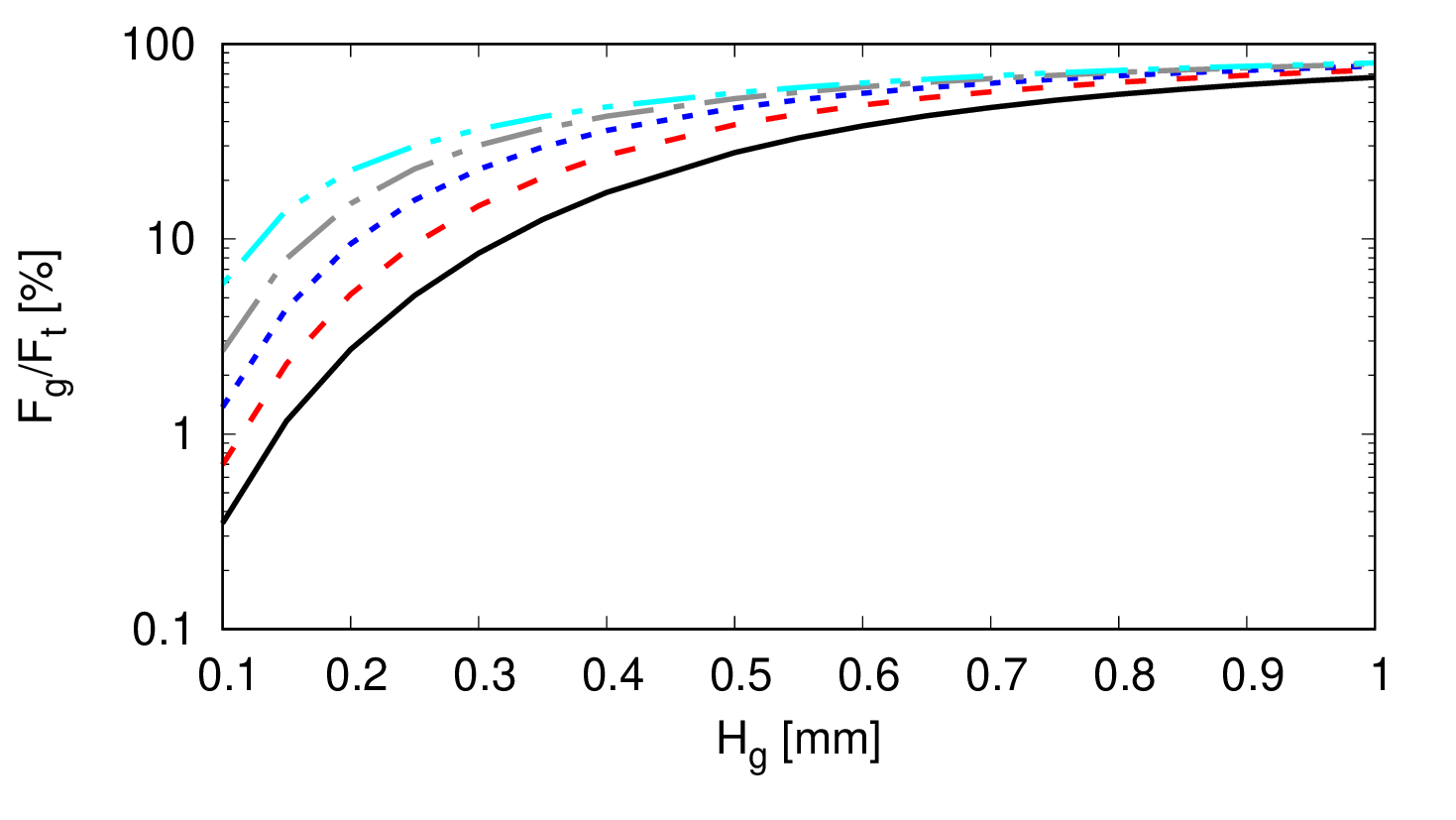} \\
Gap width $B_{\mathrm{g}}=30\, \mathrm{cm}$\\
\includegraphics[width=\factor\linewidth]{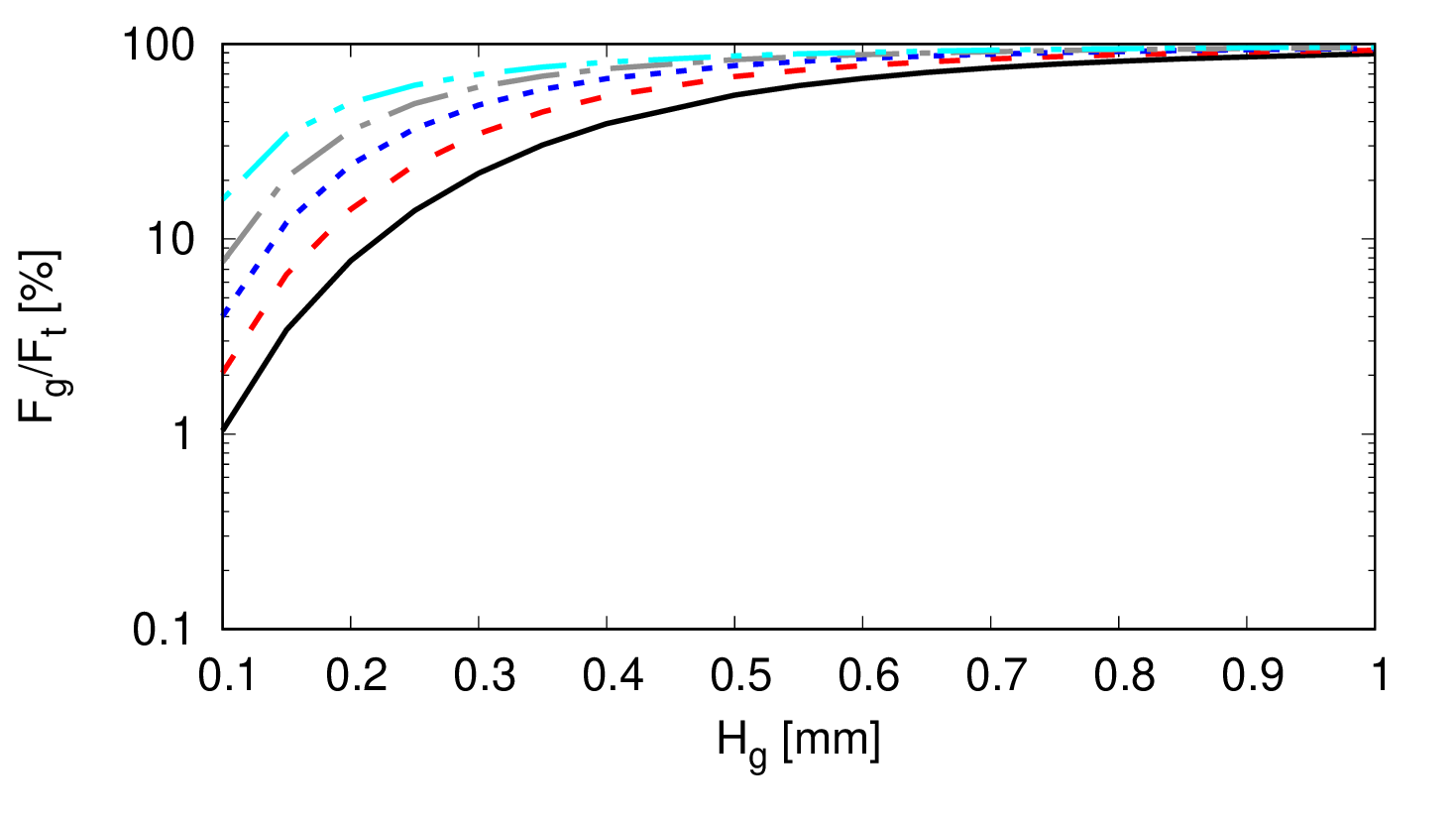} \\
\end{small}
\end{myfont}
\caption{Analytical predictions for the flow rate $F_{\mathrm{g}}$ through the gap  as percentage of the total flow rate $F_{\mathrm{t}}=30\, \mathrm{L/min}$, as a function of gap height $H_{\mathrm{g}}$, for a mask with viscous porous resistance $C_{\mathrm{m}}=2000\, \mathrm{m/s}$ and different gap widths $B_{\mathrm{g}}$ and gap lengths $L_{\mathrm{g}}$} \label{FIGFt0FgLOGtheory}
\end{center}
\end{figure}

From Figs. \ref{FIGFt0Fg}-\ref{FIGdpCm1000}, the  analytical approach from Sect. \ref{SECtheory} can be considered sufficiently validated. Therefore, Fig. \ref{FIGFt0FgLOGtheory} presents analytical predictions without backup from simulation data. The figure shows that when the gap height is small (a few multiples of $0.1\, \mathrm{mm}$), then changing the seal thickness (i.e. the gap length) from a thin seal ($L_{\mathrm{g}}=0.1\, \mathrm{cm}$) to a wide seal ($L_{\mathrm{g}}=2\, \mathrm{cm}$) can change the  gap flow rate $ F_{\mathrm{g}}$ by factor $10$ or more. 
This demonstrates that the thickness of the mask seal can have a substantial influence whether or not a mask fulfills FFP2- or FFP3-requirements.

To validate that the assumptions which were made in the analytical approach from Sect. \ref{SECtheory} and in the 2D-flow simulations from Sect. \ref{SECsetup} are valid, 3D-flow simulations were performed with a realistic human head and mask geometry (cf. Fig. \ref{FIG3Dgeom})  for three cases denoted setups A, B, and C. These differ mainly with regard to the gap shape, flow rate and flow direction, as summarized in Table \ref{TAB3Dsims}.

\begin{table}[H]
\caption{Parameters of the 3D-flow simulations in Figs. \ref{FIG3Dvelmag}-\ref{FIG3DexhaleStreamline} with realistic head and mask geometries: Average gap height $H_{\mathrm{g}}$, gap width $B_{\mathrm{g}}$, gap length $L_{\mathrm{g}}$, total flow rate $ F_{\mathrm{t}} $, gap location, shape of the gap cross-section (rectangular or circular segment), and flow direction during breathing}\label{TAB3Dsims}
\begin{footnotesize}
\begin{tabular}{cccc}
 & setup A & setup B & setup C \\ 
\hline 
$H_{\mathrm{g}}$ & $0.26\, \mathrm{mm}$ & $0.66\, \mathrm{mm}$ & $1.45\, \mathrm{mm}$ \\ 
$B_{\mathrm{g}}$ & $2.5\, \mathrm{cm}$ & $2.16\, \mathrm{cm}$ & $2.32\, \mathrm{cm}$ \\ 
$L_{\mathrm{g}}$ & $1.2\, \mathrm{cm}$ & $0.71\, \mathrm{cm}$ & $0.71\, \mathrm{cm}$ \\ 
$F_{\mathrm{t}}$ & $30\, \mathrm{\frac{L}{min}}$ & $30\, \mathrm{\frac{L}{min}}$, $95\, \mathrm{\frac{L}{min}}$  & $30\, \mathrm{\frac{L}{min}}$, $95\, \mathrm{\frac{L}{min}}$ \\ 
gap  below  & left eye & both eyes & both eyes \\ 
gap section & rectangular & rectangular & circular segment \\ 
breath & inhaling & both & exhaling \\ 
\end{tabular}
\end{footnotesize}
\end{table} 

\begin{figure}[H]
\begin{center}
\includegraphics[width=\linewidth]{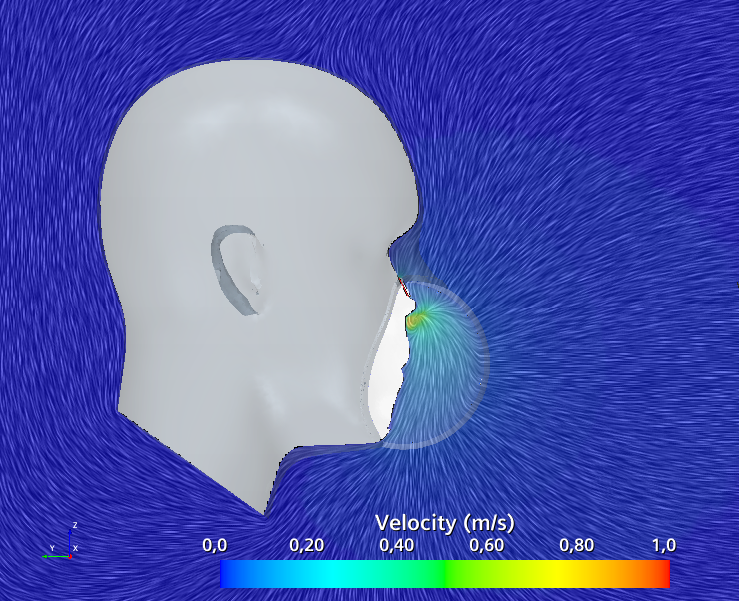}
 \includegraphics[width=\linewidth]{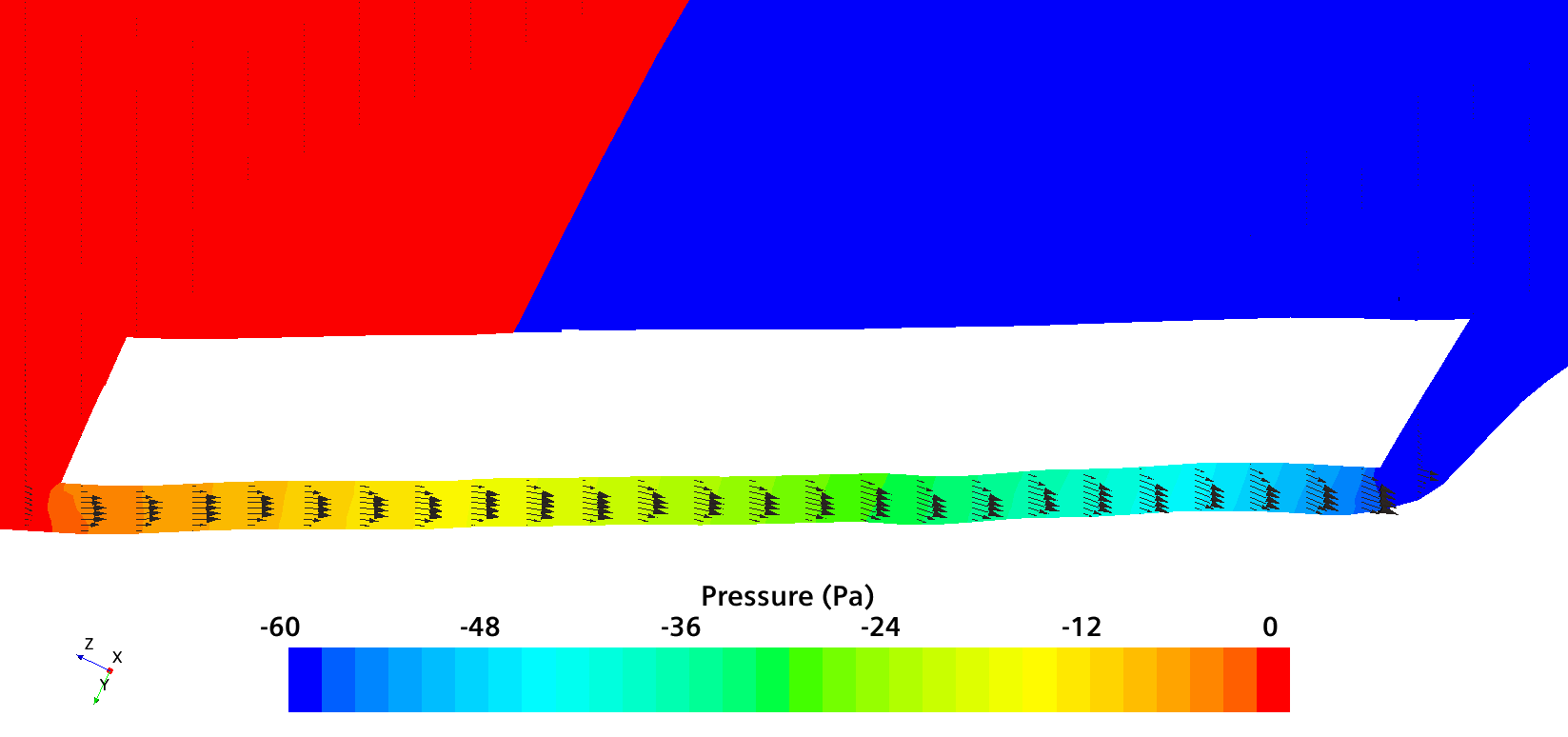}
\caption{Velocity magnitude in the plane through the mask's gap when inhaling (top) and pressure with velocity vectors for a close-up view of the gap (bottom), for 3D-flow with geometry from Fig. \ref{FIG3Dgeom} and setup A from Table \ref{TAB3Dsims}} \label{FIG3Dvelmag}
\end{center}
\end{figure}

The first 3D-flow simulation results (setup A from Table \ref{TAB3Dsims}) correspond to inhaling during normal breathing with a mask that fits tightly over the whole perimeter, except for a very small gap (average gap height $H_{\mathrm{g}}=0.26\, \mathrm{mm}$ and gap width $B_{\mathrm{g}}=2.5\, \mathrm{cm}$, cf. Figs. \ref{FIG3Dgeom} and \ref{FIG3Dvelmag}). 

\begin{figure}[H]
\begin{center}
\includegraphics[width=\factor\linewidth]{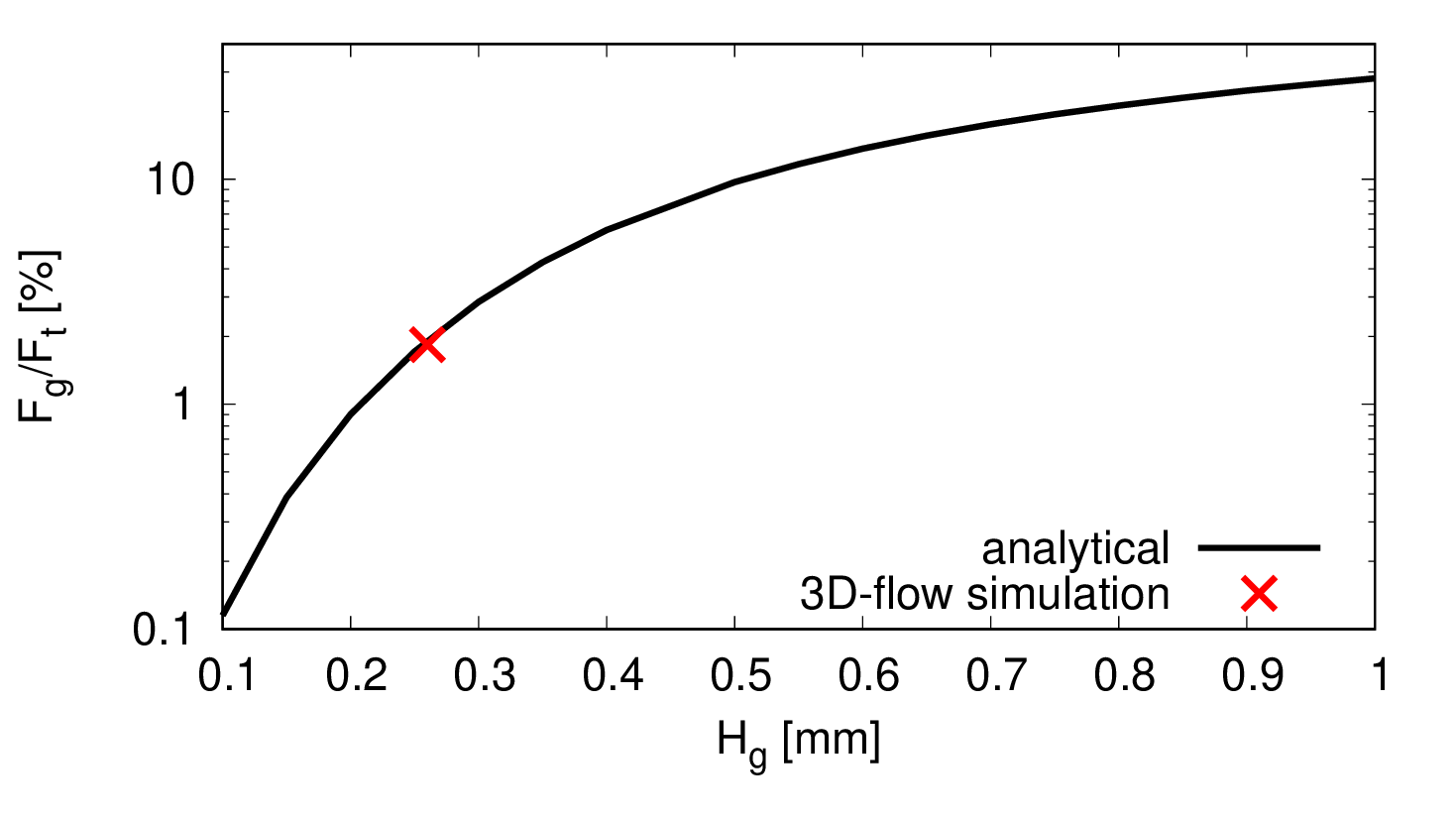}\\
\includegraphics[width=\factor\linewidth]{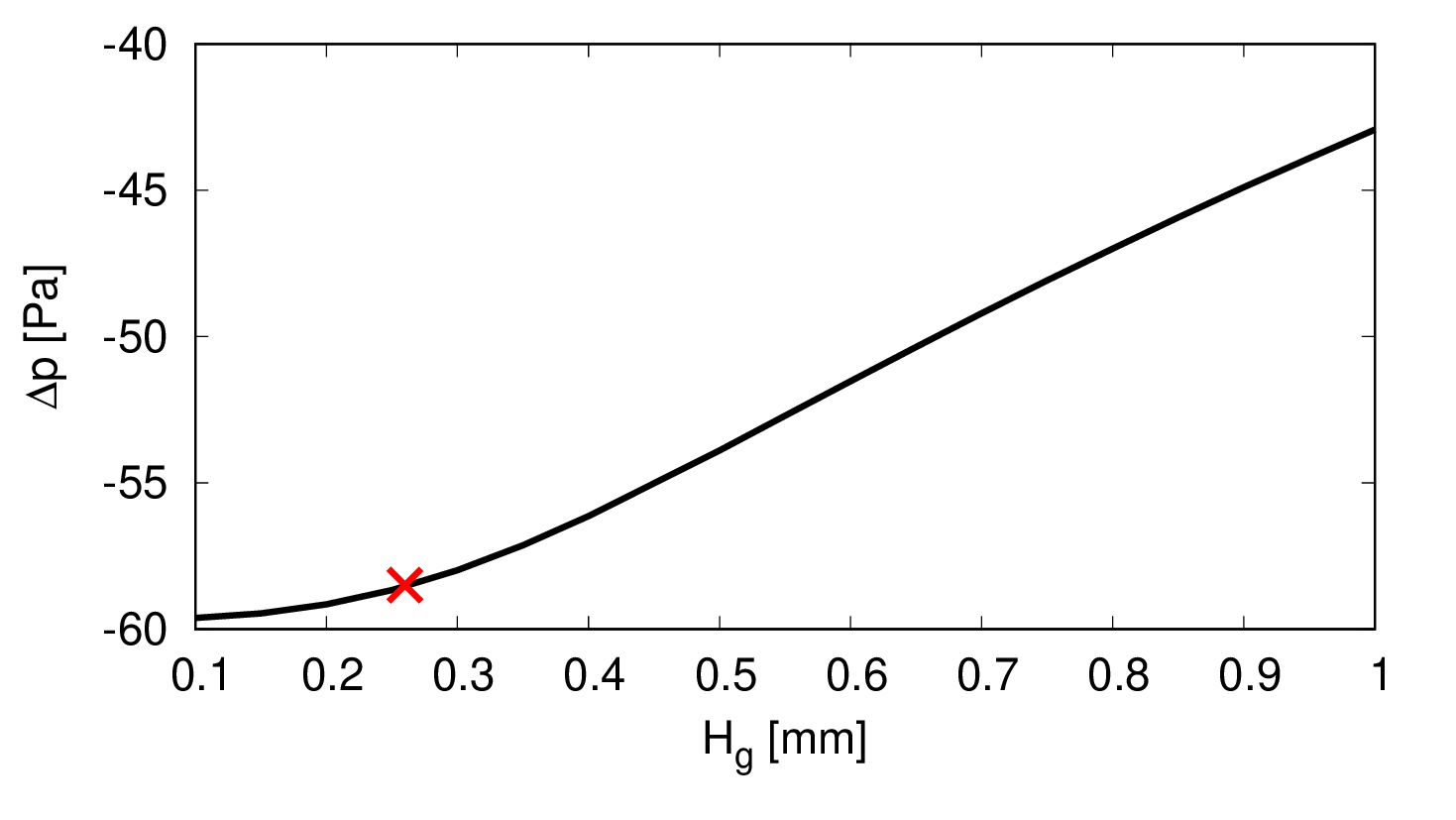}  
\caption{Analytical predictions and 3D-flow simulation results  for the flow rate $F_{\mathrm{g}}$ through the gap  as percentage of the total flow rate $F_{\mathrm{t}}=30\, \mathrm{L/min}$ (top) and  pressure drop $\Delta p$ across the filter (bottom), as a function of gap height $H_{\mathrm{g}}$; for the human head and realistic mask geometry from Fig. \ref{FIG3Dgeom} with setup A from Table \ref{TAB3Dsims}
} \label{FIGFt0Fg3D}
\end{center}
\end{figure}

Figures \ref{FIG3Dvelmag}-\ref{FIGFt0Fg3D} show that the analytical predictions for gap flow rate and pressure within the mask agree well with 3D-flow simulation results, demonstrating that previous assumptions were justified. The results show that the pressure (not shown) can indeed be approximated as constant within the mask. The largest flow velocities occur within the gap, as expected. Although the gap is very small, $1.84\%$ of the flow enters the mask through the gap, which just fulfills FFP3-requirements that leakage must be below $2\%$. Thus if there would be another such gap below the other eye, or if the seal would be slightly thinner (i.e. smaller gap length $L_{\mathrm{g}}$), FFP3-requirements would not be fulfilled.

The second 3D-flow simulation results (setup B from Table \ref{TAB3Dsims}) correspond to inhaling and exhaling during either normal breathing or deep breathing with a mask that fits tightly over the whole perimeter, except for two small gaps (average gap height $H_{\mathrm{g}}=0.66\, \mathrm{mm}$ and total gap width $B_{\mathrm{g}}=2.16\, \mathrm{cm}$). 

\begin{figure}[H]
\begin{center}
\includegraphics[width=\factor\linewidth]{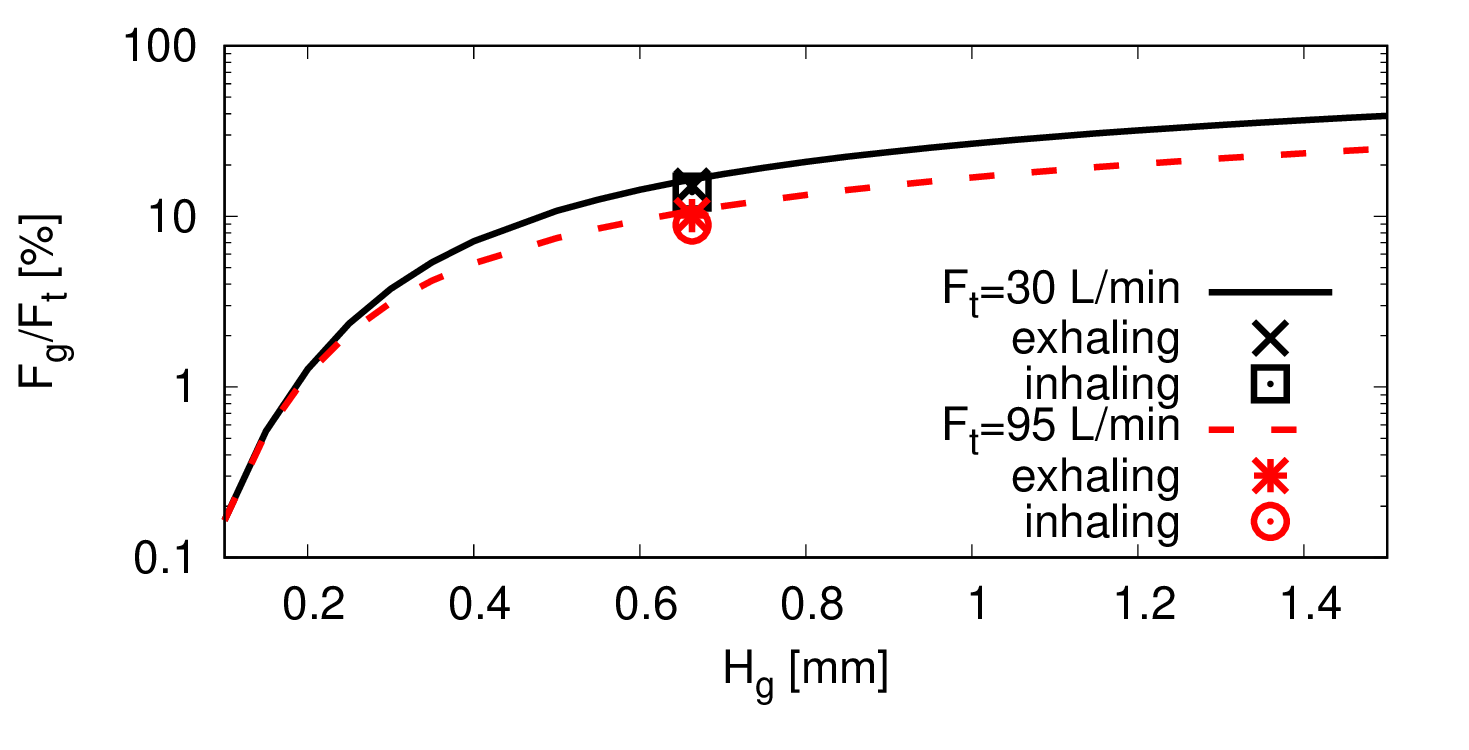} 
\includegraphics[width=\factor\linewidth]{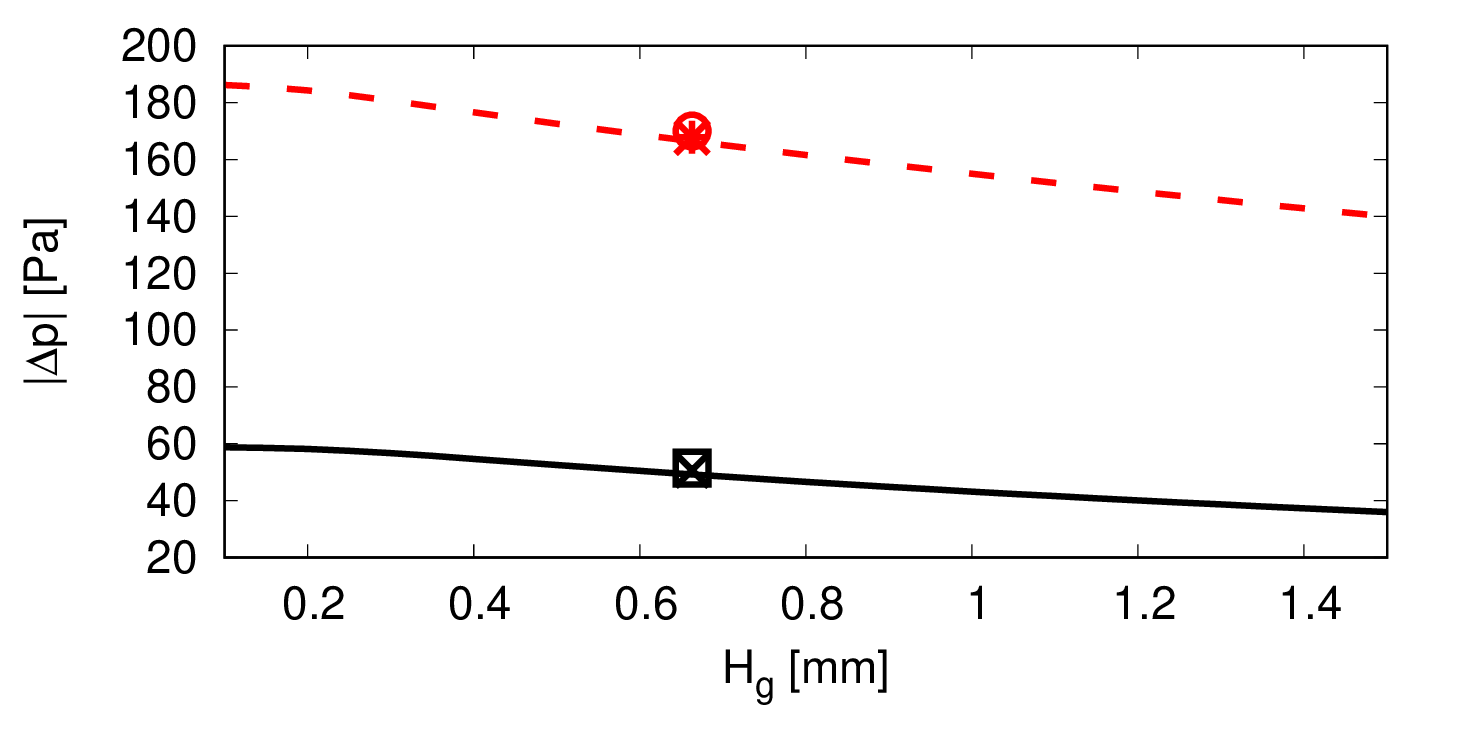} 
\caption{Analytical predictions (lines) and 3D-flow simulation results (points) for the flow rate $F_{\mathrm{g}}$ through the gap  as percentage of the total flow rate $F_{\mathrm{t}}$ (top) and absolute pressure drop $|\Delta p|$ across the filter (bottom), as a function of gap height $H_{\mathrm{g}}$; for the human head and realistic mask geometry setup B from Table \ref{TAB3Dsims}, both for inhaling and exhaling
} \label{FIGFt0Fg3Dnarrow}
\end{center}
\end{figure}

Figure \ref{FIGFt0Fg3Dnarrow} shows that, although the gaps are still comparatively small, between  $8.9\%$ (inhaling with total flow rate $F_{\mathrm{t}}=95\, \mathrm{L/min}$) and $15\%$ (exhaling with total flow rate $F_{\mathrm{t}}=30\, \mathrm{L/min}$) of the flow enters or leaves the mask through the gap unfiltered, i.e. neither FFP3- nor FFP2-requirements (i.e. less than $8\%$ leakage) are fulfilled. As before, the percentage flow through the gap is larger for lower flow rates. The gap flow rates $F_{\mathrm{g}}$ were ca. $10\%$ larger for exhaling than for inhaling, which was attributed to the different edge angles at gap inlet and outlet for this specific geometry (i.e. different resistance for entering the gap). The agreement with analytical results from the 1D model is again remarkably good.

The third 3D-flow simulation results (setup C from Table \ref{TAB3Dsims}) correspond to exhaling during either normal or deep breathing with a mask that fits tightly over the whole perimeter, except for  two medium-sized gaps (average gap height $H_{\mathrm{g}}=1.45\, \mathrm{mm}$ and total gap width $B_{\mathrm{g}}=1.45\, \mathrm{cm}$). The gaps are located at both sides of the nose, as is typical for many self-made or surgical masks. However, the selected gap size is still small compared to the range of typical gap sizes for such masks.

For exhaling with a total flow rate $F_{\mathrm{t}}=30\, \mathrm{L/min}$, ca. $29.5\%$  of the flow leaves the mask through the gap unfiltered, i.e. neither FFP3-, FFP2 nor FFP1-requirements (i.e. less than $22\%$ leakage) are fulfilled. The pressure drop was $\Delta p = 42.1\, \mathrm{Pa}$. A grid study was performed on two grids with $8.7 \cdot 10^{5}$ cells and $2.9 \cdot 10^{6}$ cells, and the results for flow rates and pressure drop differed by less than $0.25\%$, indicating that the discretization was sufficiently fine. 
The results demonstrate how sensitive the protection provided by the mask is to airflow leakage through comparatively small gaps. 

\begin{figure}[H]
\begin{center}
\includegraphics[width=0.85\linewidth]{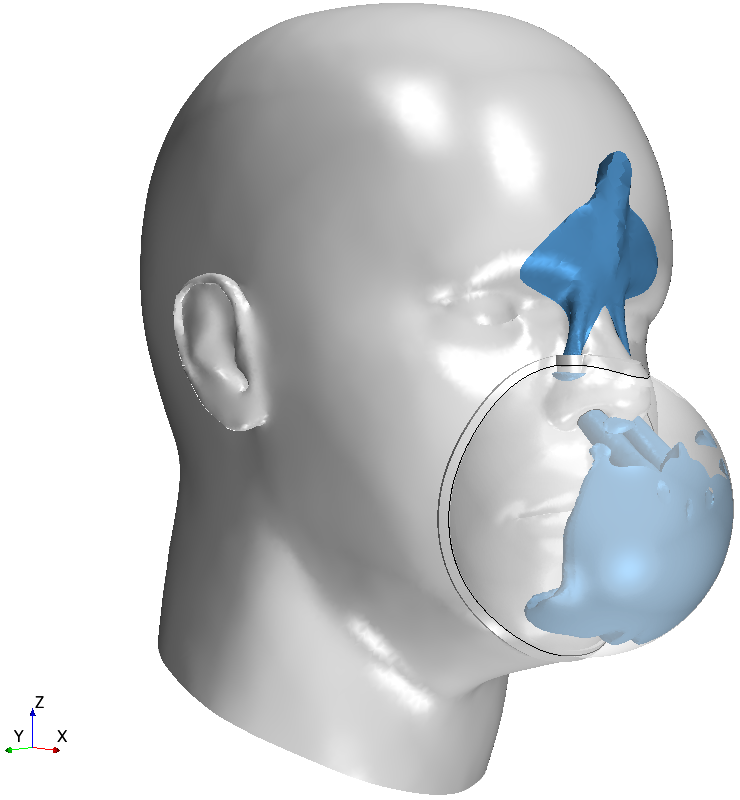} 
\includegraphics[width=\linewidth]{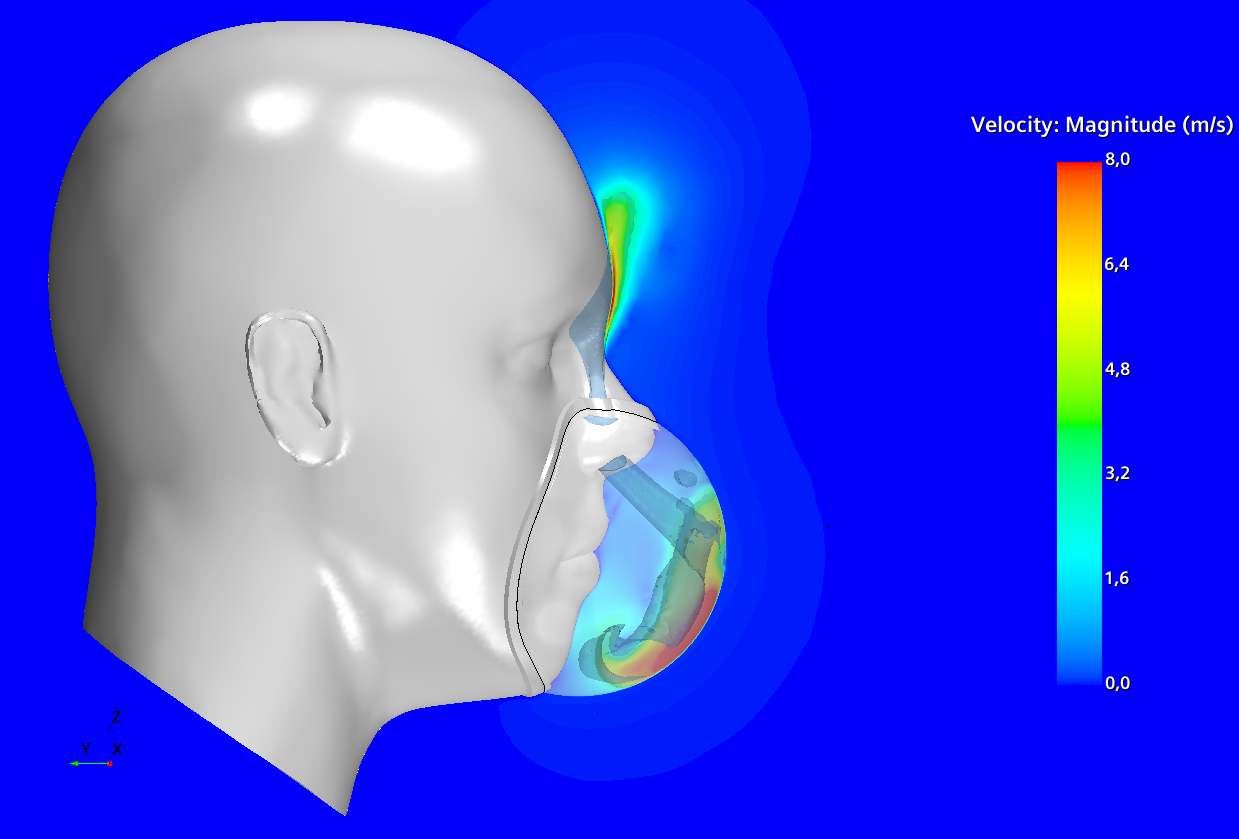} 
\caption{Isosurface where velocity magnitude $|\mathbf{u}|=3\, \mathrm{m/s}$ (blue) for exhaling with total flow rate $F_{\mathrm{t}}=95\, \mathrm{L/min}$ and mask with small gaps beneath each eye (top), and side-view including velocity magnitude  at center-plane (bottom); for setup C from Table \ref{TAB3Dsims}} \label{FIG3DexhaleIsoU}
\end{center}
\end{figure}

\begin{figure}[H]
\begin{center}
\includegraphics[width=0.9\linewidth]{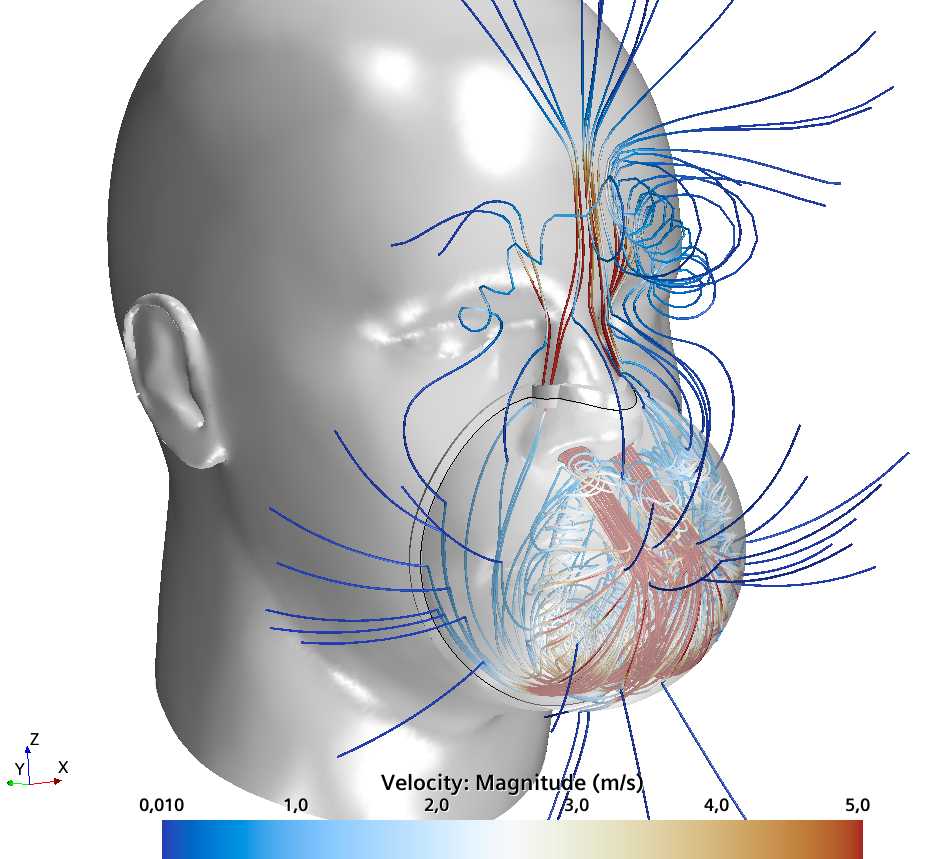} 
\includegraphics[width=\linewidth]{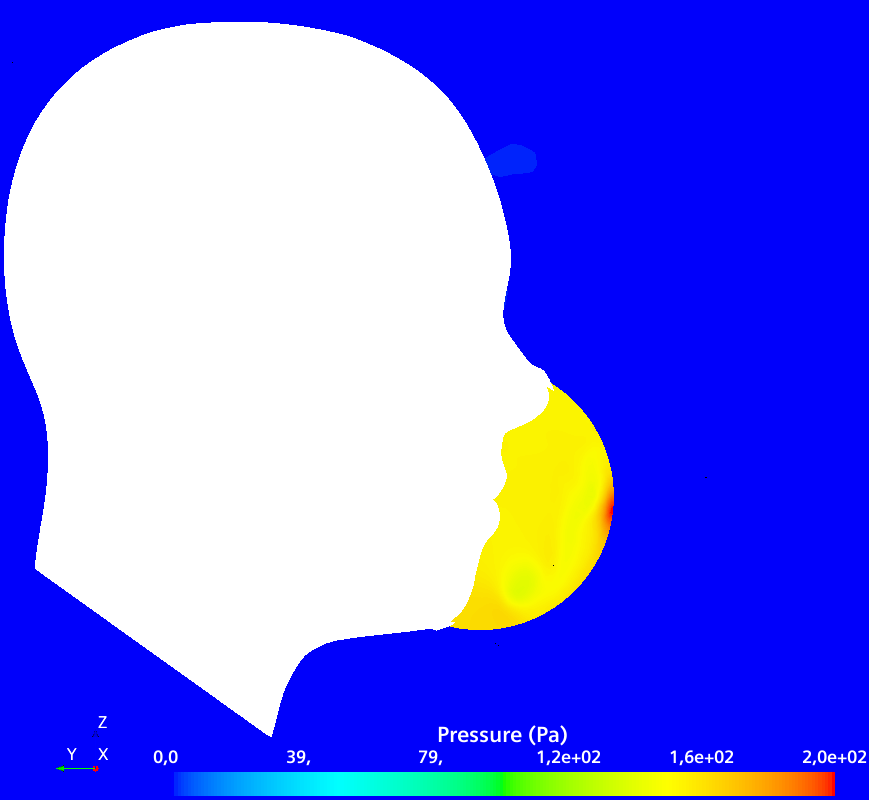} 
\caption{Streamlines colored by velocity magnitude $|\mathbf{u}|$ (top) and pressure in center-plane (bottom) corresponding to Fig. \ref{FIG3DexhaleIsoU};  where the exhaled jet hits the mask, the pressure increases locally: within the rest of the mask the pressure is roughly uniform, whereas it is almost perfectly uniform for the inhaling case} \label{FIG3DexhaleStreamline}
\end{center}
\end{figure}

Finally, Figs. \ref{FIG3DexhaleIsoU}-\ref{FIG3DexhaleStreamline} present results for exhaling of air with a total flow rate of  $F_{\mathrm{t}}=95\, \mathrm{L/min}$, where $18.5\%$ of the flow leaves the mask through the gap unfiltered with a  pressure drop of $\Delta p = 152.4\, \mathrm{Pa}$. Figure \ref{FIG3DexhaleIsoU} shows that the gap directs the airflow towards the eyes with comparatively large velocities (locally more than $8\, \mathrm{m/s}$, i.e. up to Beaufort number 5 `fresh breeze'). 
Figure \ref{FIG3DexhaleStreamline} shows how the exhaled breath from the nostrils hits the mask, is diverted sideways, partially passes through the mask with low velocities and partially is accelerated through the narrow gap and blows against the eyes and the forehead. If the mask is worn in combination with glasses, such a flow can be visualized by the fogging of the glasses. 

\FloatBarrier

\section{Discussion}
\label{SECdisc}
The present results demonstrate that tight fitting of face masks is of paramount importance. If the mask does not fit to the face perfectly, i.e. without a gap, airflow containing virus-carrying droplets  leaks into or out of the mask unfiltered. The results in Sect. \ref{SECres} demonstrate that already for gap heights $H_{\mathrm{g}} > 0.2\, \mathrm{mm}$, the total inward leakage will typically be larger than $2\%$, so that FFP3-requirements are not fulfilled anymore. For gap heights $H_{\mathrm{g}}\approx 0.4\, \mathrm{mm}$, ca. $5\%$ to $30\%$ of the air can enter the mask unfiltered through the gap, so that mostly FFP2- or even FFP1-requirements are not fulfilled, while for gaps of height $H_{\mathrm{g}}=1\, \mathrm{mm}$ more than $70\%$ of the air may pass  through the gap unfiltered. 

The gap height $H_{\mathrm{g}}$ had the largest influence on the flow rate: It was found that increasing gap height $H_{\mathrm{g}}$ by a factor of $2$ can increase the flow rate $F_{\mathrm{g}}$ through the gap by a factor of up to $10$. 
The seal thickness (i.e. gap length $L_{\mathrm{g}}$) also influenced the filtering performance substantially:
For small gap heights ($\lesssim 0.3\, \mathrm{mm}$), changing the seal thickness from a thin seal ($L_{\mathrm{g}}=0.1\, \mathrm{cm}$) to a wide seal ($L_{\mathrm{g}}=2\, \mathrm{cm}$) can change the gap flow rate $F_{\mathrm{g}}$ by factor $10$ or more. In contrast, changing the gap width $B_{\mathrm{g}}$ had a comparatively small influence, and changing the filter material had an even smaller effect. 

For the design of FFP-type masks it is therefore critical to ensure that the masks fit tightly without a gap, even if put on in a hurry by a layperson. Further,  the seal thickness should be made as large as possible, to reduce the negative effects in case that there is a gap between face and mask.

The present results confirm the experimental observation by  Lee et al. (2017) that for lower flow rates the percentage of leaked airflow increases. Moreover, the present findings indicate that the large variation in protection observed for FFP2- and FFP3-type masks in literature (with total inward leakage of $80\%$ or more for some test participants) resulted from imperfect fitting of the masks. Already for gaps which can hardly be distinguished by the naked eye, the mask may loose most of its protection. Therefore, it appears that the design of many typical FFP-type masks does not adequately consider the importance of tight fitting of the mask and may not provide reliable protection. Thus, fit-testing with suitable instruction can be recommended (cf. Lepelletier et al., 2019).

From the above discussion follows that most designs for self-made masks and surgical masks are not suitable for filtering out the majority of exhaled droplets, i.e. droplets smaller than $5\, \mathrm{\mu m}$, because they do not fit tightly and thus most air will enter and leave the mask through the gaps instead of through the filter. Only large droplets that do not follow the flow of air may be filtered more effectively. Future work could model larger droplets as Lagrangian particles, to determine via flow simulations the amount of larger droplets that pass through the gap as a function of the particle size and gap location. 

\begin{figure}[H]
\begin{center}
\includegraphics[width=\linewidth]{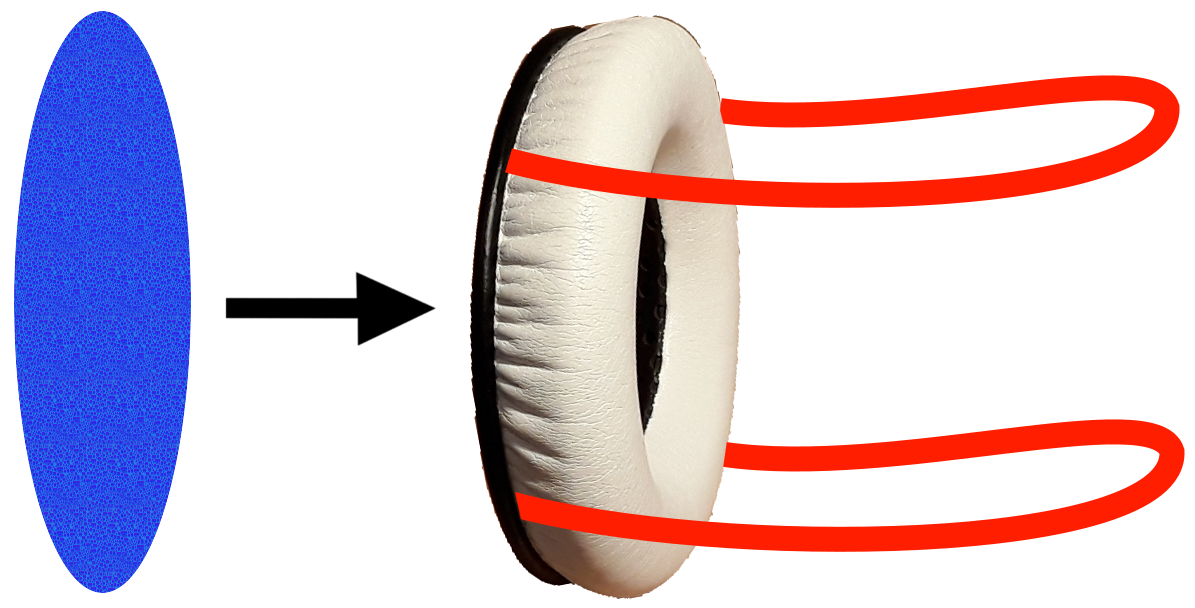}
\caption{Design suggestion for a tight-fitting FFP-type mask: a flexible headphone earpiece (white) with cushion cross-sectional diameter of ca. $2\, \mathrm{cm}$ is pressed onto the face by two adjustable bands (red) with sufficient pressure so that no gap occurs; to the other side of the earpiece, a suitable filter material of arbitrary shape is fitted tightly; in this manner a reusable mask can be constructed which provides satisfactory protection and can be comfortably worn for several hours} \label{FIGffp3maskDesignSuggestion}
\end{center}
\end{figure}

From the present findings, it can be expected that the design of most self-made masks as well as industrial masks can be improved by attaching an impermeable, sufficiently thick sealing material to the mask rim. This material should be flexible enough to allow tight fitting for different face geometries, which can be achieved by selecting a deformable material and pressing it tightly against the face by adjustable bands.  Furthermore, the material should be comfortable enough to allow wearing durations of several hours. An example for such a material are the ear-pieces of typical passive noise-cancelling headphones. The mask filter can be made of the same materials used so far. The shape of the filter is not important because the present results show that the pressure within the mask is basically uniform. The suggested improved design is illustrated in Fig. \ref{FIGffp3maskDesignSuggestion}.

\begin{figure}[H]
\begin{center}
\includegraphics[height=0.35\linewidth]{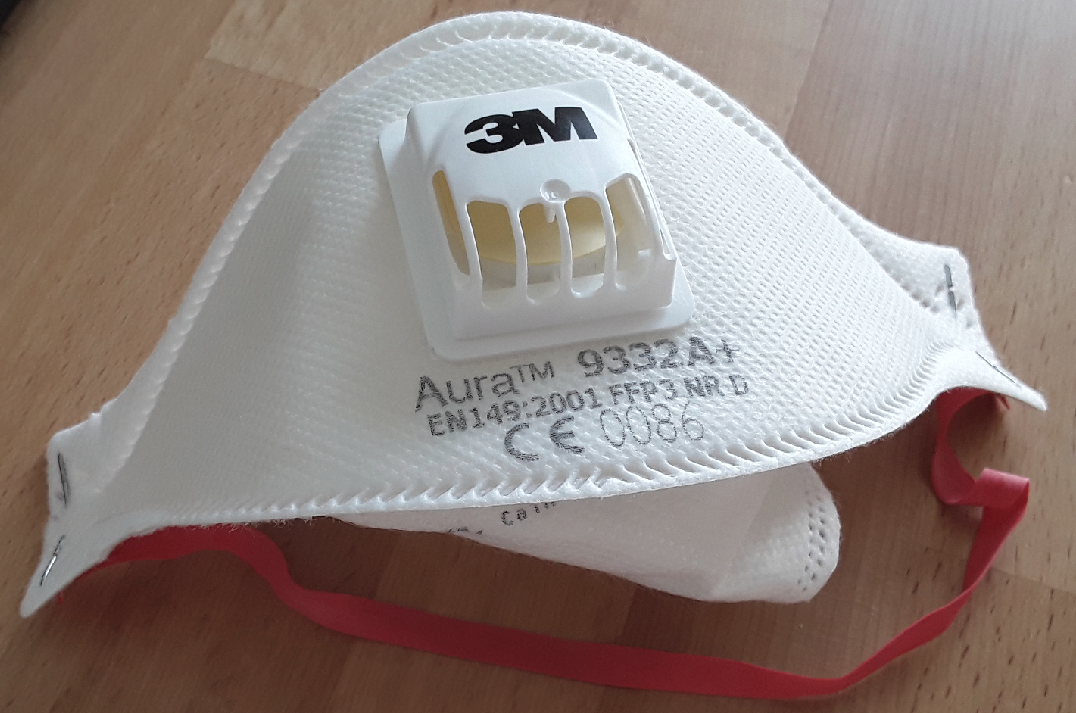} 
\includegraphics[height=0.4\linewidth]{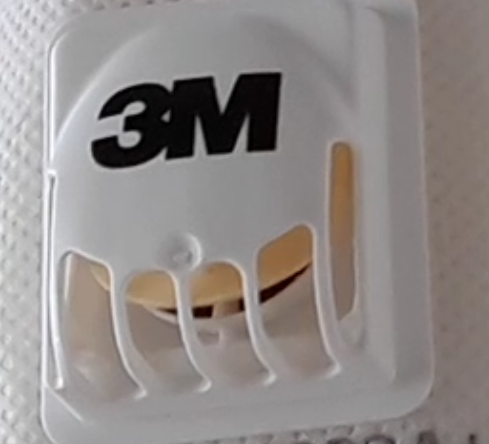}
\caption{Left: FFP3-mask with one-way valve; right: during exhaling, the one-way-valve (yellow) opens and humid air leaves the mask unfiltered} \label{FIGffp3mask}
\end{center}
\end{figure}

The present results have implications also for masks with one-way valves as shown in Fig. \ref{FIGffp3mask}. One-way valves  can reduce the breathing effort and humidity inside the mask:  When exhaling, the valve opens a gap of height $H_{\mathrm{g}} \approx 1-4\, \mathrm{mm}$ and width $B_{\mathrm{g}}\approx 2\, \mathrm{cm}$. Thus from the present results it can be expected that for these masks most air leaves the mask through the valve unfiltered. Therefore, such a mask does not provide adequate protection for others when worn by an infected person. Especially, such masks should not be worn by healthcare workers or other professions where the risk of being infected is increased.

For pandemic viruses such as COVID19, which survives at most several days on  surfaces (Kampf et al., 2020), ordinary citizens  may only require a few masks, which can be stored in a dry, warm place and after a few days can be reused, and thus would remain reusable for years. With suitable material choice, the mask can even be disinfected more quickly, for example by placing it in an oven at temperatures in the order of $80\, ^{\circ}\mathrm{C}$ for ca. half an hour (Mackenzie, 2020).  In this manner, production and distribution of FFP3-masks for the general public can be realized comparatively cost-effectively, so that social and economic consequences of a pandemic might be reduced by this measure.

\section*{Acknowledgments}
The authors are grateful to Prof. Moustafa Abdel-Maksoud, Institute for Fluiddynamics and Ship Theory, Hamburg University of Technology, Hamburg, Germany for support  and valuable discussions. The availability of the CFD software Simcenter STAR-CCM+ for this study, provided by Siemens Industry Software, is highly appreciated.

\section*{References}

\begin{hangparas}{2em}{1}
%\bibliography{jafm}

Andersen, B. M. (2019). Protection of Upper Respiratory Tract, Mouth and Eyes. In Prevention and Control of Infections in Hospitals (pp. 129-146). Springer, Cham.

Asadi, S.,  Bouvier, N., Wexler, A. S.,  and Ristenpart, W. (2020). The coronavirus pandemic and aerosols: Does COVID-19 transmit via expiratory particles?, \textit{Aerosol Science and Technology}, 54:6, 635-638, DOI: 10.1080/02786826.2020.1749229

Cherrie, J. W., Apsley, A., Cowie, H., Steinle, S., Mueller, W., Lin, C., Horwell, C. J., Sleeuwenhoek, A., and Loh, M. (2018). Effectiveness of face masks used to protect Beijing residents against particulate air pollution. \textit{Occupational and environmental medicine}, 75(6), 446-452.

Edwards, D. A., Man, J. C., Brand, P., Katstra, J. P., Sommerer, K., Stone, H. A.,   Nardell, E., and Scheuch, G. (2004). Inhaling to mitigate exhaled bioaerosols. \textit{Proceedings of the National Academy of Sciences}, 101(50), 17383-17388.

Fabian, P., McDevitt, J. J., DeHaan, W. H., Fung, R. O., Cowling, B. J., Chan, K. H., Leung, G. M., and Milton, D. K. (2008). Influenza virus in human exhaled breath: an observational study. \textit{PloS one}, 3(7).

Fairchild, C. I., and Stampfer, J. F. (1987). Particle concentration in exhaled breath. \textit{American Industrial Hygiene Association Journal}, 48(11), 948-949.

Ferziger, J. H., Perić, M., and Street, R. L. (2020). Computational methods for fluid dynamics. Berlin: Springer.

Idelchik, I. E. (1986). Handbook of hydraulic resistance. Washington, DC, Hemisphere Publishing Corp., 1986, 662 p. Translation.

Howard, J. (2020). Should you wear a mask? US health officials re-examine guidance amid coronavirus crisis, \textit{CNN online article}, March 31, 2020, link: \url{https://edition.cnn.com/2020/03/31/health/coronavirus-masks-experts-debate/index.html}

Jung, H. (2014). Performance of Filtration Efficiency, Pressure Drop and Total Inward Leakage in Anti-yellow Sand Masks, Quarantine Masks, Medical Masks, General Masks and Handkerchiefs (Doctoral dissertation, Seoul National University).

Jung, H., Kim, J., Lee, S., Lee, J., Kim, J., Tsai, P., and Yoon, C. (2014). Comparison of filtration efficiency and pressure drop in anti-yellow sand masks, quarantine masks, medical masks, general masks, and handkerchiefs. \textit{Aerosol Air Qual Res}, 14(14), 991-1002.

Kampf, G., Todt, D., Pfaender, S., and Steinmann, E. (2020). Persistence of coronaviruses on inanimate surfaces and its inactivation with biocidal agents. \textit{Journal of Hospital Infection}.

Lee, S. A., Hwang, D. C., Li, H. Y., Tsai, C. F., Chen, C. W., and Chen, J. K. (2016). Particle size-selective assessment of protection of European standard FFP respirators and surgical masks against particles-tested with human subjects. \textit{Journal of healthcare engineering}.

Lee, S. A., Chen, Y. L., Hwang, D. C., Wu, C. C., and Chen, J. K. (2017). Performance evaluation of full facepiece respirators with cartridges. \textit{Aerosol and Air Quality Research}, 17(5), 1316-1328.

Lepelletier, D., Keita-Perse, O., Parneix, P., Baron, R., Glélé, L. S. A., Grandbastien, B., and French Society for Hospital Hygiene working group. (2019). Respiratory protective equipment at work: good practices for filtering facepiece (FFP) mask. \textit{European Journal of Clinical Microbiology and Infectious Diseases}, 38(11), 2193-2195.

Mackenzie, D. (2020). Reuse of N95 Masks. \textit{Engineering} (Beijing, China).

Milton, D. K., Fabian, M. P., Cowling, B. J., Grantham, M. L., and McDevitt, J. J. (2013). Influenza virus aerosols in human exhaled breath: particle size, culturability, and effect of surgical masks. \textit{PLoS pathogens}, 9(3).

Mueller, W., Horwell, C. J., Apsley, A., Steinle, S., McPherson, S., Cherrie, J. W., and Galea, K. S. (2018). The effectiveness of respiratory protection worn by communities to protect from volcanic ash inhalation. Part I: filtration efficiency tests. \textit{International journal of hygiene and environmental health}, 221(6), 967-976.

Ntlailane, M. G. L., and Wichmann, J. (2019). Effectiveness of N95 respirators for nanoparticle exposure control (2000–2016): a systematic review and meta-analysis. \textit{Journal of Nanoparticle Research}, 21(8), 170.

Papineni, R. S., and Rosenthal, F. S. (1997). The size distribution of droplets in the exhaled breath of healthy human subjects. \textit{Journal of Aerosol Medicine}, 10(2), 105-116.

Penconek, A., Drążyk, P., and Moskal, A. (2013). Penetration of diesel exhaust particles through commercially available dust half masks. \textit{Annals of occupational hygiene}, 57(3), 360-373.

Schlichting, H., and Gersten, K. (2017). Boundary-layer theory. 9th edition, Springer-Verlag Berlin Heidelberg, DOI: 10.1007/978-3-662-52919-5.

Serfozo, N., Ondráček, J., Zíková, N., Lazaridis, M., and Ždímal, V. (2017). Size-resolved penetration of filtering materials from CE-marked filtering facepiece respirators. \textit{Aerosol Air Qual. Res}, 17, 1305-1315.

Steinle, S., Sleeuwenhoek, A., Mueller, W., Horwell, C. J., Apsley, A., Davis, A., Cherrie, J. W., and Galea, K. S. (2018). The effectiveness of respiratory protection worn by communities to protect from volcanic ash inhalation. Part II: Total inward leakage tests. \textit{International journal of hygiene and environmental health}, 221(6), 977-984.

Tellier, R. (2006). Review of aerosol transmission of influenza A virus. \textit{Emerging infectious diseases}, 12(11), 1657.

Tropea, C., and Yarin, A. L. (2007). Springer handbook of experimental fluid mechanics. Springer Science \& Business Media.

van der Sande, M., Teunis, P., and Sabel, R. (2008). Professional and home-made face masks reduce exposure to respiratory infections among the general population. \textit{PLoS One}, 3(7).

van Doremalen, N., Bushmaker, T., Morris, D. H., Holbrook, M. G., Gamble, A., Williamson, B. N., Tamin, A., Harcourt, J. L., Thornburg, N. J., Gerber, S. I., de Wit, E.,  and Lloyd-Smith, J. O. (2020). Aerosol and surface stability of SARS-CoV-2 as compared with SARS-CoV-1. \textit{New England Journal of Medicine}, 382(16), 1564-1567.

World Health Organization. (2020). 
Advice on the use of masks in the context of COVID-19: interim guidance, 6 April 2020 
(No. WHO/2019-Cov/IPC$\_$Masks/2020.3). World Health Organization.

\end{hangparas}

\end{document}